\documentclass[%
 reprint,
superscriptaddress,
longbibliography,
nobibnotes,
 amsmath,amssymb,
 aps,
 prl
]{revtex4-2}

\usepackage{amsmath}
\usepackage{amssymb}
\usepackage{amsthm}
\usepackage{physics}
\usepackage{graphicx}
\usepackage{dcolumn}
\usepackage{bm}
\usepackage{hyperref}
\usepackage{bibunits}
\defaultbibliographystyle{apsrev4-2}
\defaultbibliography{DIQKD}

\usepackage{caption}
\usepackage{subcaption}
\usepackage{mathdots}
\usepackage{mathtools}

\theoremstyle{plain}
\newtheorem{thm}{Theorem}
\newtheorem{lem}[thm]{Lemma}
\newtheorem{prop}[thm]{Proposition}
\newtheorem{cor}[thm]{Corollary}

\theoremstyle{definition}
\newtheorem{dfn}[thm]{Definition}

\theoremstyle{remark}

\usepackage{xcolor}

\begin{document}
\begin{bibunit}
\nocite{MyLongBibliography}

\preprint{APS/123-QED}

\title{Fundamental limits of multipartite entanglement and common key distribution in quantum networks}

\author{Makoto Ishihara}
 \email{llmakomako.arg1076@keio.jp}
 \affiliation{%
 Department of Electronics and Electrical Engineering, Keio University, 3-14-1 Hiyoshi, Kohoku-ku, Yokohama 223-8522, Japan
}%
\author{Wojciech Roga}
 \affiliation{%
 Department of Electronics and Electrical Engineering, Keio University, 3-14-1 Hiyoshi, Kohoku-ku, Yokohama 223-8522, Japan
}%
\author{Masahiro Takeoka}%
 \email{takeoka@elec.keio.ac.jp}
\affiliation{%
 Department of Electronics and Electrical Engineering, Keio University, 3-14-1 Hiyoshi, Kohoku-ku, Yokohama 223-8522, Japan
}%
\affiliation{
National Institute of Information and Communications Technology (NICT), Koganei, Tokyo 184-8795, Japan.
}

\date{\today}

\begin{abstract}
Distributing multipartite entanglement and establishing a common secret key are important basic tasks in quantum networks. Although the ultimate limits of these tasks have been widely investigated in bipartite settings, less is known in multipartite settings. In this work, we establish fundamental limits on the distribution of Greenberger-Horne-Zeilinger (GHZ) states and common secret keys over quantum networks. To this end, we introduce the marginal entanglement balance, an entanglement monotone that yields upper bounds on the distillation rates of GHZ states and common secret keys. We determine the asymptotic GHZ distillation rate for networks of bipartite pure states. In particular, three Bell pairs shared in a triangle network yield exactly 1.5 GHZ states per copy of the input state in the asymptotic limit. Our bound is also applicable to the problem of distributing GHZ states and/or common keys over a quantum network consisting of point-to-point quantum channels. In particular, we give a capacity for such a quantum network consisting of the erasure channels which model heralded photon loss. In addition, we investigate the effect of adding a quantum repeater in a network on multipartite entanglement distribution. Furthermore, we present fundamental limits of distillation rates in the cases where nodes aim to share not only GHZ states but also Bell states.

\end{abstract}

\maketitle


\section*{Introduction}
Quantum networks~\cite{Kimble2008The, Wehner2018Quantum} have attracted considerable interest because they enable applications such as quantum key distribution~\cite{Xu2020, Pirandola2020}, distributed quantum sensing~\cite{Zhang2021Distributed}, and distributed quantum computing~\cite{Knorzer2026Distributed}. Although many studies focus on bipartite entanglement and secret-key distribution over networks, multipartite entanglement, especially in the form of Greenberger-Horne-Zeilinger (GHZ) states~\cite{Greenberger1990Bells}, can enable a broader range of applications, including multipartite quantum key distribution (QKD)~\cite{Epping2017}, known as conference key agreement~\cite{Murta2020}, quantum secret sharing~\cite{Hillery1999Quantum}, and quantum network clocks~\cite{Komar2014A}, as well as fundamental tests of quantum mechanics~\cite{Horodecki2009Quantum, Pan2012Multiphoton}.

Practical quantum channels are subject to loss and noise, which limit the rate at which GHZ states can be distributed. Therefore, there exists a limit on a distribution rate of GHZ states over quantum networks. Then, efficient distribution schemes of GHZ states and common secret keys over networks have been widely investigated~\cite{Aschauer2005Multiparticle, Smolin2005Entanglement, Kruszynska2006Quantum, Cuquet2012Growth, Wallnofer2016Twodimensional, Pirker2018Modular, Wallnofer2019Multipartite, Vrana2019Distillation, Streltsov2020Rates, Shimizu2025Simple}. Here, a fundamental question arises: what is the ultimate limit of distributing multipartite entanglement and common secret key over a given quantum network?

The maximum achievable distribution rate of entanglement (secret key) per network round, with one use of each edge channel per round and unlimited local operations and classical communication (LOCC), is defined as the entanglement capacity (secret key capacity). For bipartite entanglement and secret key distribution, the capacity has been extensively explored for a point-to-point quantum channel~\cite{Takeoka2014Fundamental, Takeoka2014Squashed, Pirandola2017Fundamental, Wilde2017Converse}, a broadcast quantum channel~\cite{Takeoka2017Unconstrained}, and quantum networks~\cite{Azuma2016Fundamental, Pirandola2019Endtoend}. Despite recent efforts~\cite{Seshadreesan2016Bounds, Bauml2017Fundamental, Bauml2020Linear, Das2021Universal}, however, less is known for the multipartite entanglement/common key distribution capacity. This is mainly due to the little knowledge on the upper bound on distillable GHZ states/common keys, although it has long been a fundamental problem in quantum information theory~\cite{Horodecki2009Quantum, Bennett2000Exact, Linden2005}. A basic open problem is to determine the optimal rate of GHZ-state distillation from multipartite quantum states. In particular, consider a very simple tripartite state $\ket{\psi}_{ABC} = \ket{\Phi}_{AB} \otimes \ket{\Phi}_{AC} \otimes \ket{\Phi}_{BC}$, where $\ket{\Phi} = (\ket{00} + \ket{11})/\sqrt{2}$ is the Bell state. For this state, spanning-tree packing achieves a rate of $3/2$ GHZ states per input copy. However, a recent study based on multipartite Rains entanglement showed that its upper bound remains 2 for this state~\cite{Murray2026Flood}. Therefore, the optimal GHZ-distillation rate has remained undetermined.

\begin{figure}[tbp]
\centering
\includegraphics[keepaspectratio, scale=0.6]{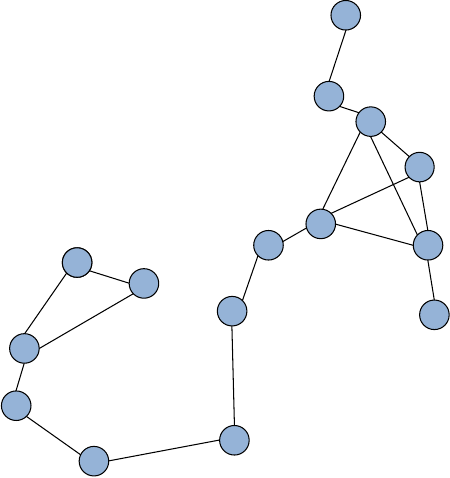}
\caption{\label{fig:NetworkSchematic}
Schematic of a quantum network. All nodes (blue circles) are connected by point-to-point quantum channels.}
\end{figure}

In this work, we answer to this question by introducing a new upper bound on the distillable GHZ state/common secret key. This bound is tight for the above simeple example, and more importantly, is applicable to establish the fundamental limit of multipartite entanglement/common secret key distribution in quantum networks.
For the latter, we consider  a quantum network, shown in Fig.~\ref{fig:NetworkSchematic}, where the nodes (parties) are connected by point-to-point quantum channels. 
All or part of the nodes aim to share multipartite entanglement or common secret key using quantum channels and unlimited LOCC among the nodes.


Our findings are summarized as follows. 
First, we introduce a new entanglement monotone called marginal entanglement balance (MEB). With this monotone, we derive upper bounds on the distillable GHZ and common-key rates. This bound itself gives substantial progress on the longstanding problem in multipartite entanglement theory, that is, it determines the exact GHZ distillation rate for a certain class of multipartite quantum states, including the above tripartite state.

Second, we consider the GHZ state/common secret key distribution over all nodes in a given quantum network. By using the MEB, we can establish a fundamental upper bound on the GHZ state/common key distribution capacity in a quantum network where the nodes are connected by teleportation-simulable channels. In particular, we determine the GHZ/common key distribution capacity for an erasure-channel network.

Third, we consider how the `network quantum repeater' can overcome the capacity of the repeaterless quantum network. We show an example of such a quantum repeater and also derive a fundamental bound on the GHZ state/common key distribution rate in a network with the repeater. In addition, we show that all the above results are applicable to common secret key distribution in a trusted-node-based network, which has been deployed in field trials globally~\cite{Peev2009The, Sasaki2011Field, Stucki2011Longterm, Wang2014Field}.

Finally, we generalize the above scenario to the following one: simultaneous distillation of GHZ states and Bell states in a given quantum network. Although some applications in a quantum network may prefer GHZ states to achieve high performance, there exist cases where distributing Bell states is enough to complete the applications. Therefore, determining achievable simultaneous distillation rates of GHZ and Bell states plays an important role in evaluating the ultimate performance of practical quantum networks. Using the MEB, we can establish fundamental limits of the distillation rates of GHZ and Bell states in this scenario.

\begin{figure}[tbp]
\centering
\includegraphics[keepaspectratio, scale=0.47]{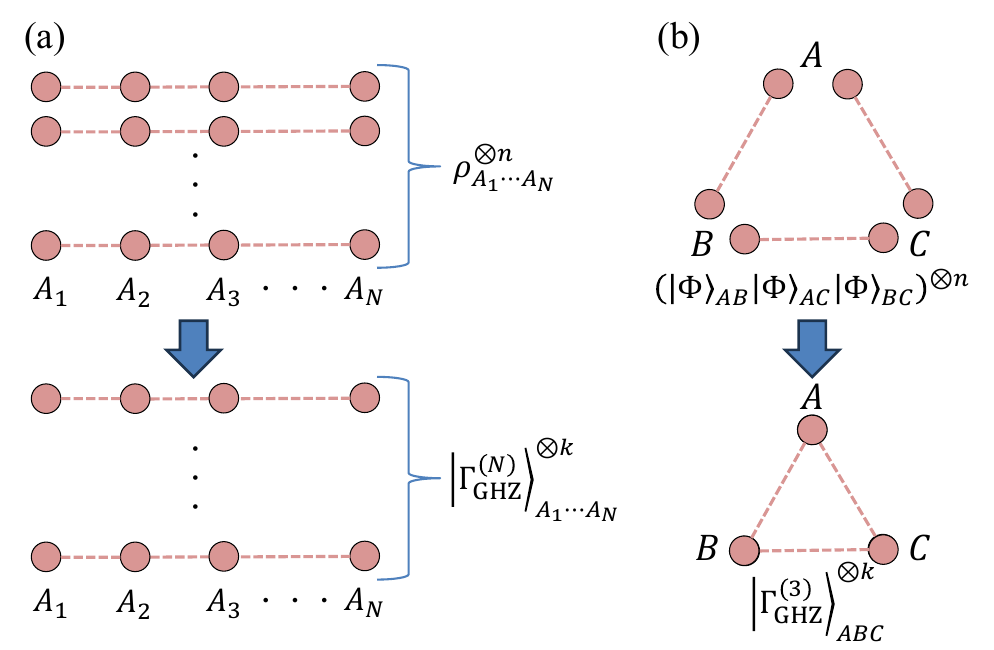}
\caption{\label{fig:DistillationSchematic}
Schematic of GHZ-state distillation from $n$ copies of (a) a general $N$-partite quantum state $\rho$ and (b) a product of three Bell states $\ket{\Phi}$ shared by three parties, $A$, $B$, and $C$.}
\end{figure}

\section*{RESULTS}
\subsection*{GHZ distillation from shared bipartite states}
Before going to quantum networks, we start with a simpler problem where the $N$ distant parties, $A_1, \ldots, A_N$, aim to distill $k$ copies of the GHZ state $\ket{\Gamma_\text{GHZ}^{(N)}} = (\ket{0\cdots 0} + \ket{1 \cdots 1})/\sqrt{2}$ from $n$ copies of an $N$-partite quantum state $\rho$ initially shared by the parties via LOCC (Fig.~\ref{fig:DistillationSchematic} (a) and (b)). The distillable GHZ rate $D_{\text{GHZ}, N}(\rho)$ is defined as the maximal achievable rate $r = k / n$ at $n \rightarrow \infty$ (see Supplementary Information A for a rigorous definition). As mentioned above, the distillable GHZ rate is not known even for a simple tripartite state $\ket{\Phi}_{AB}\ket{\Phi}_{AC}\ket{\Phi}_{BC}$ (Fig.~\ref{fig:DistillationSchematic}(b)). From two copies of the input state, the parties can construct three edge-disjoint spanning trees and obtain one GHZ state from each tree. This achieves a rate of 3/2. However, the currently known upper bounds on the rate~\cite{Horodecki2009Quantum, Das2021Universal, Yang2009Squashed, Augusiak2009Multipartite, Murray2026Flood} are not better than the entropy of entanglement of a bipartitioned system, which evaluates to two in this case. Therefore, there remains a significant gap.

To close this gap, we develop an upper bound on the distillable GHZ rate with the new entanglement monotone MEB:

\begin{thm}\label{thm:MEBupper}
    The distillable GHZ rate of an $N$-partite finite-dimensional quantum state $\rho$ is upper bounded as
    \begin{equation}
        D_{\text{GHZ}, N}(\rho) \leq \tilde{M}_{N, \infty}(\rho),
    \end{equation}
    where $\tilde{M}_{N, \infty}$ represents the $N$-partite MEB.
\end{thm}

See Supplementary Information B for the formal definition of the MEB and the proof of Theorem~\ref{thm:MEBupper}. Note that the MEB defined for pure states and its convex-roof extension are LOCC monotones. The quantity $\tilde{M}_{N, \infty}$ appearing in the above theorem is a tensor-power-additive upper bound on the convex-roof extension. In addition, note that the above bound is a weak converse bound, i.e., it holds in the asymptotic limit of many copies of $\rho$ and in the limit as the distillation error tends to zero. 

The above upper bound is applicable to arbitrary finite-dimensional $N$-partite quantum states. However, it is particularly useful for a quantum state which is a product of bipartite states. Consider a finite graph on $N$ vertices $G = (V, E)$ where $V$ represents a set of vertices, one vertex corresponds to one party, $E$ represents a set of edges connecting the parties, and $|V| = N$. Suppose a quantum state $\rho_e$ is shared on an edge $e \in E$. The state of the whole system becomes $\bigotimes_e \rho_e$. Let $E_D(\rho_e)$ and $E_C(\rho_e)$ denote distillable entanglement and entanglement cost for $\rho_e$. Then, the distillable GHZ rate is bounded as follows:

\begin{thm}\label{thm:ProductBounds}
    The distillable GHZ rate of an $N$-partite finite graph with a state $\bigotimes_e \rho_e$ is bounded as
    \begin{equation}\label{eq:MixedNetwork}
    \begin{split}
        \min_{\mathcal{P} : |\mathcal{P} |\geq 2} \frac{\sum_{e \in E(\mathcal{P})} E_D(\rho_e)}{|\mathcal{P}|-1} &\leq D_{\text{GHZ}, N}(\bigotimes_e \rho_e)\\
        &\leq \min_{\mathcal{P}:|\mathcal{P}|\geq 2} \frac{\sum_{e \in E(\mathcal{P})} E_C(\rho_e)}{|\mathcal{P}|-1},
    \end{split}
    \end{equation}
    where $\mathcal{P}$ denotes the partitioning of the parties and $E(\mathcal{P})$ denotes a set of edges connecting different blocks. In particular, when $E_C (\rho_e) = E_D (\rho_e)$, we have
    \begin{equation}
        D_{\text{GHZ}, N}(\bigotimes_e \rho_e)
        = \min_{\mathcal{P}:|\mathcal{P}|\geq 2} \frac{\sum_{e \in E(\mathcal{P})} E_D(\rho_e)}{|\mathcal{P}|-1}.
    \end{equation}
\end{thm}

\begin{proof}
    Here we sketch the proof of Theorem~\ref{thm:ProductBounds}. See Supplementary Information C for details.

    We first consider the upper bound on the distillable GHZ. To obtain this bound, we consider a Bell-state graph where, for each edge $e \in E$, there is a $d_e$-dimensional maximally entangled state $\ket{\Phi_{d_e}} \coloneqq \frac{1}{\sqrt{d_e}} \sum_{j = 0}^{d_e-1} \ket{jj}$. The whole system of such a Bell-state graph can be expressed as $\ket{\Phi_G} = \bigotimes_{e \in E} \ket{\Phi_{d_e}}$. Then, we find that the MEB of this state becomes 
    \begin{equation}\label{eq:MEBonBellNetworkPartition}
        M_{N, \infty} (\Phi_G) = \min_{\mathcal{P}:|\mathcal{P}|\geq2} \frac{\sum_{e \in E(\mathcal{P})} \log_2 d_e}{|\mathcal{P}|-1}.
    \end{equation}
    When there are many copies of the maximally entangled state on edge $e$, we can prepare $n$ copies of $\rho_e$ using approximately $n E_C(\rho_e)$ Bell states. Therefore, combining Eq.~\eqref{eq:MEBonBellNetworkPartition} with Theorem~\ref{thm:MEBupper}, we derive the upper bound on the distillable GHZ rate.

    The lower bound is obtained as follows. First, for each edge $e$, we generate the maximally entangled state from many copies of the mixed state $\rho_e$ at rate $E_D(\rho_e)$. Then, we use an LOCC protocol consisting of generating a GHZ state from pairs of the maximally entangled state via local CNOT gates, partial measurements, and the conditional bit flip~\cite{Meignant2019Distributing}. To generate a GHZ state shared among all the parties, we have to choose a set of edges connecting all the parties. Furthermore, these connections should be carefully chosen to maximize the number of generated GHZ states. This optimization problem is known as the fractional spanning-tree packing problem in graph theory. Using the celebrated theorem by Tutte~\cite{Tutte1961On} and Nash-Williams~\cite{NashWilliams1961Edge}, we obtain the lower bound in Eq.~\eqref{eq:MixedNetwork} as the maximal achievable GHZ-state-generation rate.
\end{proof}

As a consequence of Theorem~\ref{thm:ProductBounds}, we consider a case where the states $\rho_e$ on edge $e$ are all pure. Suppose edge $e = \{ A_i, A_j \}$ connects two parties $A_i$ and $A_j$. Then, we find that the distillable GHZ rate is given by
\begin{equation}
    D_{\text{GHZ}, N}(\bigotimes_e \rho_e) = \min_{\mathcal{P}: |\mathcal{P}|\geq 2} \frac{\sum_{e \in E(\mathcal{P})} s_e}{|\mathcal{P}|-1},
\end{equation}
where $s_e \coloneqq S(\text{Tr}_{A_j}[\rho_e])$ denotes the von Neumann entropy of the reduced system of $A_i$ for the state $\rho_e$. Now, we can answer the question that arose in the introduction of this paper. Namely, consider the following state
\begin{equation}
    \ket{\psi}_{ABC} = \ket{\Phi}_{AB}^{\otimes a} \ket{\Phi}_{AC}^{\otimes b}\ket{\Phi}_{BC}^{\otimes c}.
\end{equation}
The distillable GHZ rate on this state becomes
\begin{equation}
\begin{split}
    D_{\text{GHZ}, 3}(\psi) = \min\{\frac{a+b+c}{2}, a+b, a+c, b+c \},
\end{split}
\end{equation}
as discussed in detail in Supplementary Information D. When $a = b = c = 1$, we obtain $D_{\text{GHZ}, 3}(\psi) = 1.5$ as expected.

\subsection{GHZ distribution capacity for quantum networks}
We apply the above results to a quantum communication scenario, that is, GHZ distribution in the quantum network illustrated in Fig.~\ref{fig:NetworkSchematic}. Consider an $N$-partite finite graph $G = (V, E)$ and each edge $e \in E$ corresponds to a channel $\mathcal{N}_e$. Here, we consider a quantum network which is teleportation simulable, i.e., consists of teleportation-simulable channels~\cite{Bennett1996Mixed, Niset2009No, MullerHermes2012Transposition}. This class of channels includes practically important channels, such as Pauli, dephasing and erasure channels. An operation of a teleportation-simulable channel $\mathcal{N}_e$ can be perfectly simulated with a bipartite resource state $\omega_e$ and LOCC $\Lambda_e$, i.e., for a quantum state $\rho$, $\mathcal{N}_e(\rho) = \Lambda_e (\rho \otimes \omega_e)$. In other words, using a teleportation-simulable channel $n$ times can be replaced with $n$ copies of a resource state with LOCC. Therefore, the GHZ distribution capacity $Q^\leftrightarrow_{\text{GHZ}, N}$ of the teleportation-simulable quantum network $\{ \mathcal{N}_e\}$ is upper bounded by $D_{\text{GHZ}, N} (\bigotimes_{e \in E} \omega_e)$. On the other hand, when we assume that each channel is resource-seizable, i.e., we can obtain one copy of the resource state from one use of the channel and LOCC without preshared entanglement~\cite{Wilde2018Entanglement}, we can lower bound the capacity with the same quantity as the upper bound, i.e., the capacity becomes
\begin{equation}
    Q^\leftrightarrow_{\text{GHZ}, N} (\{ \mathcal{N}_e\}) = D_{\text{GHZ}, N} (\bigotimes_{e \in E} \omega_e).
\end{equation}
Then, using Theorem~\ref{thm:ProductBounds}, we have the following fundamental bound on $Q^\leftrightarrow_{\text{GHZ}, N}$:

\begin{thm}\label{thm:Teleportation}
    Consider a quantum network consisting of teleportation-simulable channels $\{ \mathcal{N}_e \}$ with resource states $\{\omega_e  \}$. Suppose that each resource state $\omega_e$ can be obtained from one use of $\mathcal{N}_e$ and LOCC without preshared entanglement. Then, the GHZ distribution capacity is bounded as follows:
    \begin{equation}\label{eq:thmTeleportation}
        \begin{split}
            \min_{\mathcal{P}: |\mathcal{P}|\geq 2} \frac{\sum_{e \in E(\mathcal{P})} E_D (\omega_e)}{|\mathcal{P}|-1} &\leq Q_{\text{GHZ}, N}^\leftrightarrow (\{ \mathcal{N}_e \})\\
            &\leq \min_{\mathcal{P}: |\mathcal{P}|\geq 2} \frac{\sum_{e \in E(\mathcal{P})} E_C(\omega_e)}{|\mathcal{P}|-1}.
        \end{split}
    \end{equation}
    In particular, when $E_C(\omega_e) = E_D (\omega_e)$, we have
    \begin{equation}
        Q_{\text{GHZ}, N}^\leftrightarrow(\{ \mathcal{N}_e \}) = \min_{\mathcal{P}: |\mathcal{P}|\geq 2} \frac{\sum_{e \in E(\mathcal{P})} E_D(\omega_e)}{|\mathcal{P}|-1}.
    \end{equation}
\end{thm}

It is known that several teleportation-simulable channels satisfy the condition $E_C(\omega_e) = E_D(\omega_e)$ for their corresponding resource states. The erasure channel is a particularly relevant example~\cite{Bennett1997Capacities}, which provides a simple model of photon loss in quantum communication. In particular, an erasure channel describes the transmission of a quantum state where the receiver can identify an erasure from an orthogonal flag. For the qubit erasure channel, the distillable entanglement and the entanglement cost of the corresponding resource state are known to be $E_D(\omega_e) = E_C(\omega_e) = 1-p_e$ where $p_e$ represents the erasure probability~\cite{Pirandola2017Fundamental, Wilde2018Entanglement}. Therefore, together with Theorem~\ref{thm:Teleportation}, we have the following fundamental limit of GHZ-state distribution over optical quantum networks:

\begin{thm}\label{thm:erasure}
    The GHZ distribution capacity of the quantum network $\{ \mathcal{N}_e^{p_e}\}$ consisting of point-to-point qubit erasure channels is given by
    \begin{equation}
        Q^\leftrightarrow_{\text{GHZ}, N}(\{ \mathcal{N}_e^{p_e} \}) = \min_{\mathcal{P}:|\mathcal{P}|\geq 2} \frac{\sum_{e \in E(\mathcal{P})} (1-p_e)}{|\mathcal{P}|-1},
    \end{equation}
    where $\mathcal{N}_e^{p_e}$ represents an erasure channel on edge $e$ with erasure probability $p_e$.
\end{thm}

The upper bound in Eq.~\eqref{eq:thmTeleportation} can be extended to more general case. Suppose, in the $N$-partite teleportation-simulable quantum network, GHZ states are distributed over only part of the parties, $A_1, \ldots, A_S$, where $S < N$. Let $\bar{\mathcal{P}}$ be a partition of all $N$ parties into at least two blocks, each containing at least one of the $S$ target parties. Then, we have the following upper bound:

\begin{thm}\label{thm:Partial}
    The GHZ distribution capacity over $S$ parties $Q^\leftrightarrow_{\text{GHZ}, S} (\{ \mathcal{N}_e \})$ is bounded by
    \begin{equation}
        Q_{\text{GHZ}, S}^\leftrightarrow (\{ \mathcal{N}_e \}) \leq \min_{\bar{\mathcal{P}}:|\bar{\mathcal{P}}| \geq 2} \frac{\sum_{e \in E(\bar{\mathcal{P}})} E_C(\omega_e)}{|\bar{\mathcal{P}}|-1}.
    \end{equation}
\end{thm}

See Supplementary Information E for the proof.

\subsection{Common key distribution}
All of the above theorems hold not only for GHZ-state distillation but also for common secret key distillation for multiple parties by replacing $D_{\text{GHZ}, N}$ and $Q_{\text{GHZ}, N}^\leftrightarrow$ with $K_{\text{c}, N}$ and $P_{\text{c}, N}^\leftrightarrow$, respectively, where $K_{\text{c}, N}$ is the distillable common-key rate and $P_{\text{c}, N}^\leftrightarrow$ is the common private capacity. Recently, upper bounds on common-key distillation rates in pair-entangled networks have been derived under local measurements with shared randomness followed by classical postprocessing~\cite{Neumann2026Bounds}. In contrast, our MEB-based upper bound applies to general LOCC protocols and also provides bounds on GHZ-state distillation. Furthermore, our above theorems can be applied to the trusted-node-based QKD network. Therefore, our results provide fundamental limits of common key distribution in the field-deployed QKD networks~\cite{Peev2009The, Sasaki2011Field, Stucki2011Longterm, Wang2014Field}. See Supplementary Information F for the proofs. Note that spanning-tree packing has also been shown to be optimal for conference-key generation in trusted-node QKD networks where pairwise secret keys have already been established~\cite{Trushechkin2026Spanning}. Our results are consistent with this finding and further establish bounds on GHZ-state and common-key distribution directly from shared quantum states.

\begin{figure}[tbp]
\centering
\includegraphics[keepaspectratio, scale=0.4]{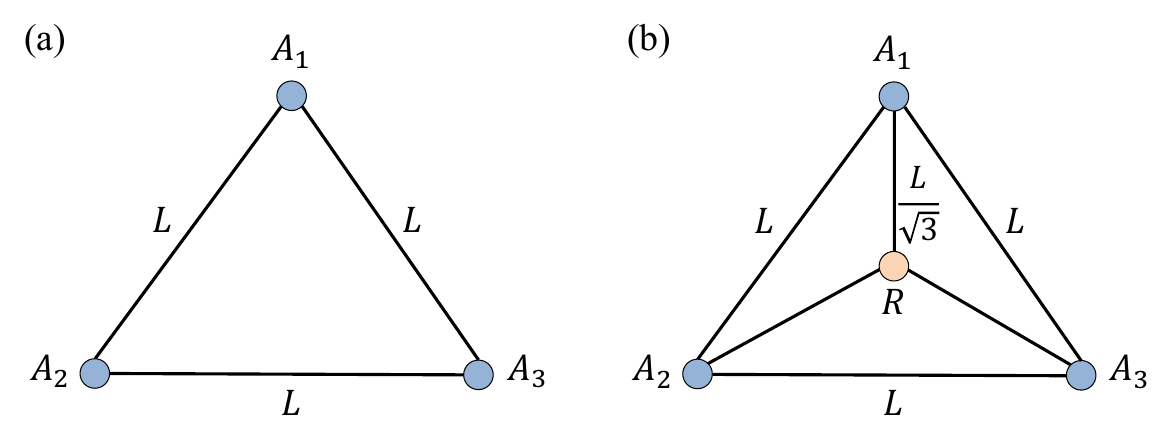}
\caption{\label{fig:ErasureNetwork}
(a) Triangle quantum network with three parties, $A_1, A_2$, and $A_3$, connected via erasure channels with distance $L$. (b) The same quantum network topology where the network quantum repeater node $R$ is located in the center. The distance of channels between $R$ and each party is $L/\sqrt{3}$.}
\end{figure}

\subsection{Network quantum repeater}
We now show how a network quantum repeater can overcome the fundamental limit (Theorem~\ref{thm:erasure}) of a repeaterless network consisting of erasure channels. Analogous to bipartite entanglement distribution, quantum repeaters can overcome the repeaterless capacity for the GHZ-state distribution in a quantum network. We call such a device a network quantum repeater. 

To this end, we first consider GHZ-state distillation in an $N$-partite complete graph $K_N$ where each pair of parties shares Bell states at a rate of $a$ per round. We add a network quantum repeater $R$ which is directly connected to all the $N$ parties, forming a complete graph on $N+1$ vertices. Suppose each party shares Bell states with the repeater $R$ at a rate of $b$ per round. The maximal GHZ-state distillation rates without and with the network repeater can be expressed as follows:

\begin{thm}\label{thm:repeater}
    The maximal GHZ-state distillation rate in an $N$-partite complete graph $K_N$ without a network repeater is
    \begin{equation}
        D_{\text{GHZ}, N}^\text{no-repeater} = \frac{N}{2}a.
    \end{equation}
    On the other hand, the GHZ-state distillation rate with a network repeater is
    \begin{equation}
        D_{\text{GHZ}, N}^\text{repeater} = \frac{N}{2}a + b.
    \end{equation}
\end{thm}

See Supplementary Information G for the proof. Note that, even with the network repeater $R$, GHZ states are distributed among $N$ parties except $R$.

\begin{figure}[tbp]
\centering
\includegraphics[keepaspectratio, scale=0.6]{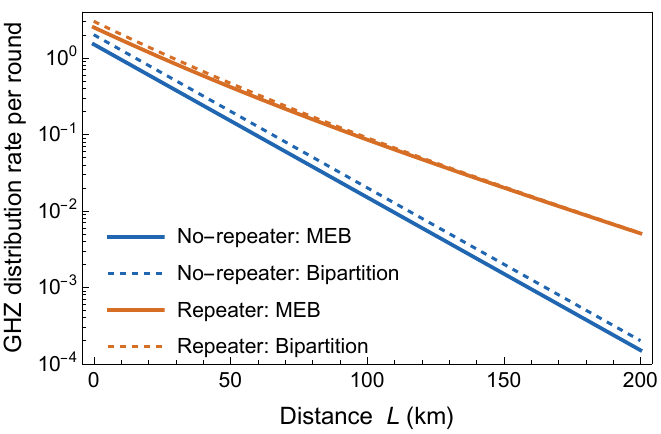}
\caption{\label{fig:ErasureCapacity}
Exact GHZ distillation rates and bipartition upper bounds without and with a network quantum repeater in the triangle erasure-channel network. Blue and orange solid (dashed) lines correspond to the MEB (bipartition) bounds without and with the repeater, respectively. We use $\alpha = 0.2$ dB/km.}
\end{figure}

With this theorem, we consider the simple example, shown in Fig.~\ref{fig:ErasureNetwork}(a), which is a triangle quantum network with three parties, $A_1, A_2$, and $A_3$. Each pair of the parties is connected by an erasure channel with transmission probability $\eta (L) = 10^{-\alpha L/10}$, where $L$ represents a channel distance and $\alpha$ represents a channel attenuation coefficient. In this case, $a = \eta(L)$ copies of the Bell state are shared between each pair of the parties per round. Here, we add a network quantum repeater $R$ and consider that the repeater is connected with the parties with erasure channels as shown in Fig.~\ref{fig:ErasureNetwork}(b). The distance between the repeater node and each party is $L/\sqrt{3}$. Therefore, for the channels between the network repeater and the parties, $b = \eta(L/\sqrt{3})$ copies of the Bell states are shared per round. Consequently, using Theorem~\ref{thm:repeater}, the GHZ-state distillation rate without and with the network repeater are expressed as $D_{\text{GHZ}, 3}^\text{no-repeater} = \frac{3}{2} \eta(L)$ and $D_{\text{GHZ}, 3}^{\text{repeater}} = \frac{3}{2} \eta(L) + \eta(L/\sqrt{3})$, respectively. In Fig.~\ref{fig:ErasureCapacity}, we plot these rates against the distance $L$ in kilometres with $\alpha = 0.2$ dB/km. We can see the effect of the network repeater, overcoming the repeaterless distillation rate. In addition, we also plot upper bounds on GHZ-state distillation rate without and with a network quantum repeater obtained by considering bipartitions of the parties (see Supplementary Information G for details). Whereas such upper bounds overestimate the maximal distillation rates, our MEB bounds provide the tight bounds on the distillation rates. Theorem~\ref{thm:repeater} also holds for common-key distillation by replacing the distillable GHZ rates with the corresponding distillable common-key rates. Consequently, the results shown in Fig.~\ref{fig:ErasureCapacity} also characterize the common private capacities of the networks without and with a network quantum repeater.

\subsection{Simultaneous distillation of GHZ and Bell states}
Finally, we consider another important problem in quantum networks, i.e., simultaneous distillation of GHZ and Bell states. Whereas some applications in quantum networks require GHZ states, Bell states are also useful, in particular, for applications between two parties in the networks. The MEB also allows us to derive fundamental limits of distillation rates in this scenario.

We first focus on a simple setting. Consider an $N$-partite complete graph $K_N$ where each pair of the parties shares the Bell state $\ket{\Phi}$. A quantum state of the whole system can be expressed as $\Phi_{K_N} = \bigotimes_{1 \leq i < j \leq N} \Phi_{A_i A_j}$. By performing an LOCC protocol, the parties aim to share the following state from $n$ copies of the input state
\begin{equation}
    T_n = (\Gamma_2^{(N)})^{\otimes m_n} \otimes \bigotimes_{1 \leq i < j \leq N} \Phi_{A_i A_j}^{\otimes k_n},
\end{equation}
where we assume that each pair of parties shares the same number of copies of the Bell state for simplicity. The corresponding distillation rates of GHZ and Bell states are given by, respectively, $g = \lim m_n/n$ and $r = \lim k_n/n$ as $n \rightarrow \infty$. Using the MEB allows us to have the following fundamental bound on the maximal achievable distillation rates (see Supplementary Information H for the proof):

\begin{thm}\label{thm:Complete}
    A nonnegative rate pair $(g, r)$ in the complete graph $K_N$ is achievable if and only if it satisfies
    \begin{equation}\label{eq:thmComplete}
        g + \frac{N}{2}r \leq \frac{N}{2}.
    \end{equation}
\end{thm}

The above example provides the simple bound on the simultaneous distillation rates of GHZ and Bell states. However, in practical quantum networks, the number of required Bell states may be different for each pair of parties. Therefore, we next consider a different setting, i.e., the triangle network shown in Fig.~\ref{fig:DistillationSchematic}(b). Suppose that three parties, $A$, $B$, and $C$, share the following input state $\psi_{ABC} = \Phi_{AB}^{\otimes a} \otimes \Phi_{AC}^{\otimes b} \otimes \Phi_{BC}^{\otimes c}$. From $n$ copies of this state, by performing LOCC, the parties aim to share the following state
\begin{equation}
    T_n = (\Gamma_2^{(3)})^{\otimes m_n} \otimes \Phi_{AB}^{\otimes k_{AB}, n} \otimes \Phi_{AC}^{\otimes k_{AC, n}} \otimes \Phi_{BC}^{\otimes k_{BC, n}}.
\end{equation}
The corresponding distillation rates are $g = \lim m_n/n$, $x = \lim k_{AB, n}/n$, $y = \lim k_{AC, n}/n$, and $z = \lim k_{BC, n}/n$ as $n \rightarrow \infty$. By using the MEB, we can also determine the tight upper bounds on these distillation rates:

\begin{thm}\label{thm:Asymmetric}
    A nonnegative rate tuple $(g, x, y, z)$ in the triangle quantum network is achievable if and only if it satisfies
    \begin{equation}\label{eq:thmAsymmetric}
        \begin{split}
            2g + x + y + z &\leq a + b + c,\\
            g + x + y &\leq a + b,\\
            g + x + z &\leq a + c,\\
            g + y + z &\leq b + c.
        \end{split}
    \end{equation}
    In particular, for fixed nonnegative Bell-state distillation rates, $x, y$, and $z$, for which a tuple $(0, x, y, z)$ satisfies Eq.~\eqref{eq:thmAsymmetric}, the optimal GHZ-state distillation rate $g_\text{max}$ becomes
    \begin{equation}
        \begin{split}
            &g_\text{max}(x, y, z) = \min\{ \frac{a+b+c-x-y-z}{2}, \\
            &a+b-x-y, a+c-x-z, b+c-y-z\}.
        \end{split}
    \end{equation}
\end{thm}

See Supplementary Information H for the proof. 

\begin{figure}[tbp]
\centering
\includegraphics[keepaspectratio, scale=0.75]{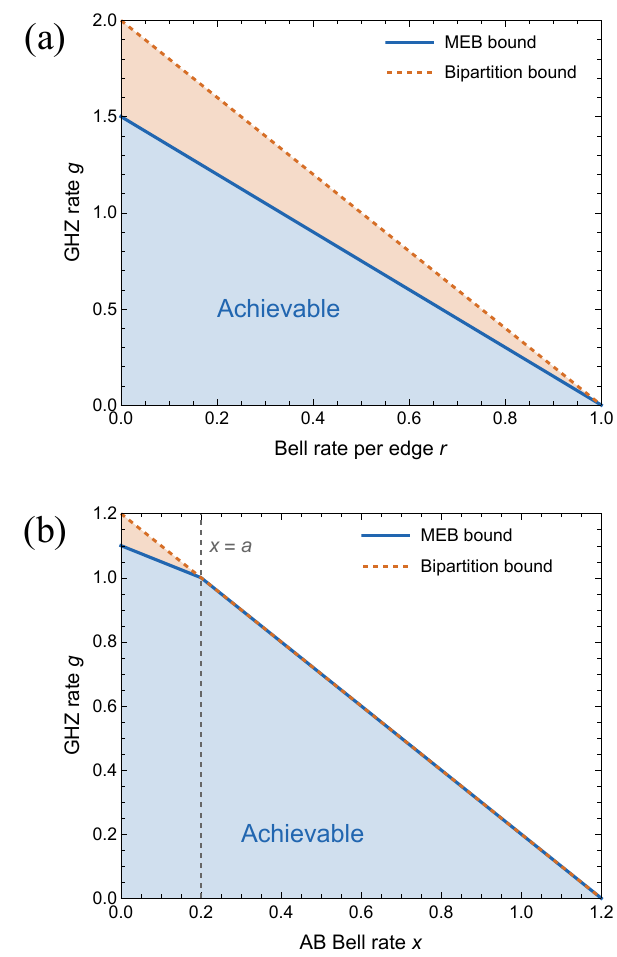}
\caption{\label{fig:SimultaneousDistributions}
Achievable GHZ and Bell-state distillation rates in a triangle quantum network for (a) equal Bell rates and (b) Bell rate between the parties $A$ and $B$ with $y = z = 0$.}
\end{figure}

We plot the achievable GHZ-state distillation rate $g$ against the Bell-state distillation rate $r$ described by Eq.~\eqref{eq:thmComplete} in Fig.~\ref{fig:SimultaneousDistributions}(a) for $N = 3$. We also show an upper bound on the distillation rates obtained with the bipartition of the three parties (see Supplementary Information H for details). Moreover, in Fig.~\ref{fig:SimultaneousDistributions}(b), we plot the achievable GHZ-state distillation rate $g$ described by Eq.~\eqref{eq:thmAsymmetric} and an upper bound with the bipartition against the Bell-state distillation rate between $A$ and $B$. Here, we use $a = 0.2$, $b = c = 1$, and $y = z = 0$. While the bipartition bounds overestimate the simultaneous distillation rates, our MEB perfectly characterizes the achievable distillation rates. These results highlight that the MEB is useful to determine the ultimate performance of practical applications in quantum networks.

\section*{Discussion}
In this paper, we establish the fundamental limit of GHZ state/common key distribution in a quantum network without and with a network quantum repeater. In particular, we determine the exact GHZ state/common key distribution capacity of a quantum network consisting of point-to-point erasure channels, corresponding to a practical optical quantum network. To derive these results, we introduce a new entanglement monotone called marginal entanglement balance which upper bounds the distillable GHZ state/common key. The MEB itself provides significant progress on the longstanding problem of determining the distillable GHZ state. The result is also applicable to establish the fundamental limit of distributing common keys over the trusted-node-based QKD networks which have been deployed in field trials. Furthermore, we establish the fundamental limit of simultaneous distillation of GHZ and Bell states in quantum networks. Therefore, our results provide the fundamental benchmark for designing and implementing future quantum networks as well as the current trusted-node-based QKD networks.

\section*{Acknowledgments}
This work was supported by JST SPRING, Grant No. JPMJSP2123, JST Moonshot R\&D, Grant No. JPMJMS256E, JST CRONOS, Grant No. JPMJCS24N6, JST COI-NEXT, Grant No. JPMJPF2221 and JST ASPIRE, Grant No. JPMJAP2427. The authors used OpenAI's ChatGPT-6.1 Sol to assist with exploratory literature searches, checks of mathematical derivations, and language editing. All AI-assisted content and calculations were independently checked by the authors.

\textbf{Note added.}---While finalizing this manuscript, we became aware of an independent work by Sun~\cite{Sun2026The}, which determines the GHZ and common-key distillation rates for networks of shared GHZ states.
These two works employ differen mathematical approaches. 
Our approach is based on a general entanglement monotone, marginal entanglement balance, and also provides bounds for mixed states and quantum channel networks, as well as a characterization of the achievable rates of simultaneous GHZ- and Bell-state distillation. Also, note that some of the authors had recognized the importance of this problem long before the present work~\cite{takeoka2020multipartiteentanglementsecretkey}.





\putbib
\end{bibunit}



\clearpage
\onecolumngrid

\begin{center}
\textbf{\large Supplementary Information}

\vspace{0.5em}

\textbf{Fundamental limits of multipartite
entanglement and common key distribution
in quantum networks}

\vspace{0.5em}

Makoto Ishihara, Wojciech Roga, and Masahiro Takeoka
\end{center}

\begin{bibunit}

\makeatletter
\def\@extra@b@citeb{@supp}
\def\@extra@binfo{@supp}
\makeatother

\nocite{MyLongBibliography}

\setcounter{equation}{0}
\renewcommand{\theequation}{S\arabic{equation}}
\renewcommand{\theHequation}{S\arabic{equation}}

\section{A. Preliminaries}
\subsection{1. LOCC-assisted GHZ distillation protocol}

We describe a rigorous definition of the distillable GHZ rate here. Suppose that $N$ parties, $A_1, \ldots, A_N$, share $n$ copies of an $N$-partite quantum state $\rho^{\otimes n}_{A_1 \cdots A_N}$. From this state with unlimited LOCC, the parties aim to share an $N$-partite GHZ state with local dimension $d$:
\begin{equation}\label{eq:GammadN}
    \ket{\Gamma_d^{(N)}} \coloneqq \frac{1}{\sqrt{d}} \sum_{j = 0}^{d-1} \ket{j \cdots j}_{A_1 \cdots A_N}, \quad \Gamma_d^{(N)} \coloneqq \ketbra{\Gamma_d^{(N)}}{\Gamma_d^{(N)}}.
\end{equation}

Let $\Lambda_n$ be an LOCC protocol of block length $n$ and let $\omega_n = \Lambda_n (\rho^{\otimes n})$. A rate $R$ is achievable if there exist positive integers $d_n $ and LOCC channels $\Lambda_n$ such that
\begin{equation}
    T(\omega_n, \Gamma_{d_n}^{(N)}) \rightarrow 0,
\end{equation}
and
\begin{equation}
    \liminf_{n \rightarrow \infty} \frac{\log_2 d_n}{n} \geq R,
\end{equation}
where $T(\rho, \sigma) \coloneqq \frac{1}{2}||\rho - \sigma||_1$ denotes the trace distance between two quantum states $\rho$ and $\sigma$. The distillable GHZ rate is the supremum of all achievable rates, that is,
\begin{equation}\label{eq:DistillableGHZ}
    D_{\text{GHZ}, N} (\rho) \coloneqq \sup \{ R: R \text{ is achievable} \}.
\end{equation}
Note that, for $d = 2^m$, $\Gamma_d^{(N)}$ is equivalent to $m$ copies of $\Gamma_2^{(N)}$ up to local unitaries. In addition, we will use the following variational representation for trace distance:
\begin{equation}\label{eq:TraceDistanceVariational}
    T(\rho, \sigma) = \max_{M} |\text{Tr}[M(\rho-\sigma)]|,
\end{equation}
where the maximization is over positive operator-valued measure (POVM) elements $M$ satisfying $0 \leq M \leq I$ with the identity $I$.

\subsection{2. Relative entropy of entanglement (REE)}
Here, we describe the definition and basic properties of relative entropy of entanglement (REE). Define the von Neumann entropy of a quantum state $\rho$ in system $X$ as $S(X)_\rho \coloneqq - \text{Tr} [\rho_X \log_2 \rho_X]$. Also, define the relative entropy between two quantum states $\rho$ and $\sigma$ as $D(\rho || \sigma) \coloneqq \text{Tr} [\rho (\log_2 \rho - \log_2 \sigma)]$. When $\text{supp} \rho \nsubseteq \text{supp} \sigma$, this becomes $+ \infty$. 

For an $N$-partite system $A_1, \ldots, A_N$, we define a set of fully separable states $\text{SEP} (A_1 : \cdots : A_N) \coloneqq \{ \sum_k p_k \rho_{A_1}^k \otimes \rho_{A_2}^k \otimes \cdots \otimes \rho_{A_N}^k\}$ where $\rho_{A_1}^k, \ldots, \rho_{A_N}^k$ are quantum states on systems $A_1, \ldots, A_N$, respectively, $p_k \geq 0$, and $\sum_k p_k = 1$. Then, we define the relative entropy of entanglement (REE) as
\begin{equation}\label{eq:REE}
    E_R(A_1: \cdots : A_N)_\rho \coloneqq \inf_{\sigma \in \text{SEP} (A_1: \cdots : A_N)} D(\rho_{A_1 \cdots A_N} || \sigma).
\end{equation}
When $N = 1$, $E_R$ vanishes. Its regularization is defined as
\begin{equation}\label{eq:ReguralizedREE}
    E_R^\infty (A_1: \cdots : A_N)_\rho \coloneqq \lim_{n \rightarrow \infty} \frac{1}{n} E_R (A_1^n : \cdots : A_N^n)_{\rho^{\otimes n}}.
\end{equation}

The REE and the regularized REE have the following basic properties:

\begin{lem}[Basic properties of REE]\label{lem:REEproperties}
For arbitrary quantum state $\rho$, the REE is nonnegative:
\begin{equation}\label{eq:REENonnegativity}
    E_R(\rho) \geq 0.
\end{equation}
The REE is a convex function, i.e.,
\begin{equation}\label{eq:REEConvexity}
    E_R (\sum_x p_x \rho_x ) \leq \sum_x p_x E_R(\rho_x).
\end{equation}
The REE is an LOCC monotone, i.e., for LOCC $\Lambda$,
\begin{equation}\label{eq:REELOCCmonotone}
    E_R (\Lambda (\rho)) \leq E_R (\rho).
\end{equation}
Consequently, the regularized REE is also an LOCC monotone
\begin{equation}\label{eq:RegularizedREELOCCmonotone}
    E_R^\infty (\Lambda (\rho)) \leq E_R^\infty (\rho).
\end{equation}
For two quantum states $\rho$ and $\omega$, the REE is subadditive, i.e.,
\begin{equation}\label{eq:REESubadditivity}
    E_R (\rho \otimes \omega) \leq E_R (\rho) + E_R (\omega).
\end{equation}
The regularized REE is weakly additive, that is,
\begin{equation}\label{eq:RegularizedREEWeakAdditivity}
    E_R^\infty(\rho^{\otimes k} ) = k E_R^\infty (\rho).
\end{equation}
\end{lem}

\begin{proof}[Proof of Lemma~\ref{lem:REEproperties}.]
Since the quantum relative entropy is nonnegative~\cite{NielsenChuang2010}, the REE is also nonnegative. Let $\sigma_x$ denote an optimal fully-separable state for $\rho_x$. Then, $\sigma = \sum_x p_x \sigma_x \in \text{SEP}$. The definition of the REE gives
\begin{equation}
    \begin{split}
        E_R(\sum_x p_x \rho_x) &\leq D(\sum_x p_x \rho_x || \sum_x p_x \sigma_x) \leq \sum_x p_x D(\rho_x ||\sigma_x)= \sum_x p_x E_R(\rho_x),
    \end{split}
\end{equation}
where the second inequality follows from the joint convexity of the quantum relative entropy~\cite{NielsenChuang2010}.

For arbitrary fully-separable state $\sigma$, $\Lambda (\sigma)$ is also a fully-separable state. Therefore,
\begin{equation}
    \begin{split}
        E_R(\Lambda (\rho)) &\leq D(\Lambda (\rho) || \Lambda (\sigma))\leq D(\rho || \sigma),
    \end{split}
\end{equation}
where the last inequality follows from the data-processing inequality~\cite{Wilde2017QIT}. Taking the infimum of the right-hand side proves the LOCC monotonicity of the REE.

Let $\sigma_\rho$ and $\sigma_\omega$ denote optimal fully-separable states corresponding to $\rho$ and $\omega$, respectively. Since $\sigma_\rho \otimes \sigma_\omega$ is a fully-separable state,
\begin{equation}
    \begin{split}
        E_R(\rho \otimes \omega) &\leq D(\rho \otimes \omega || \sigma_\rho \otimes \sigma_\omega)= D(\rho || \sigma_\rho) + D(\omega || \sigma_\omega)= E_R(\rho) + E_R(\omega),
    \end{split}
\end{equation}
where the first equality follows from the additivity of the relative entropy~\cite{Wilde2017QIT}.

Finally, by the definition of the regularized REE, we have
\begin{equation}
    \begin{split}
        E_R^\infty (\rho^{\otimes k}) &= \lim_{n \rightarrow \infty} \frac{1}{n} E_R(\rho^{\otimes kn})= k \lim_{n \rightarrow \infty} \frac{1}{kn} E_R (\rho^{\otimes kn})= k E_R^\infty (\rho).
    \end{split}
\end{equation}
\end{proof}

Since both the REE and the regularized REE are LOCC monotone, both are invariant under local isometries. In addition, since the REE is subadditive, the regularized REE is well-defined by Fekete's lemma. Furthermore, the regularized REE is convex:

\begin{lem}[Convexity of the regularized REE]\label{lem:RegularizedREEConvexity}
The regularized REE is a convex function, i.e.,
\begin{equation}\label{eq:RegularizedREEConvexity}
    E_R^\infty (\sum_x p_x \rho_x) \leq \sum_x p_x E_R^\infty(\rho_x).
\end{equation}
\end{lem}

\begin{proof}[Proof of Lemma~\ref{lem:RegularizedREEConvexity}]
    Define $e_x \coloneqq E_R^\infty(\rho_x)$. By the definition of the regularized REE, $e_x = \lim_{n \rightarrow \infty} \frac{1}{n} E_R(\rho_x^{\otimes n})$. Then, for arbitrary $\delta > 0$, a constant $C_x$, and sufficiently large $n$, we have $E_R(\rho_x^{\otimes n}) \leq n (e_x + \delta) + C_x$. Here, define $\rho^{\otimes k} \coloneqq \sum_{x^k} p_{x^k} \rho_{x_1} \otimes \cdots \otimes \rho_{x_k}$ and $p_{x^k} \coloneqq \prod_{j = 1}^k p_{x_j}$. Let $N_x (x^k)$ denote the number of occurrences of $x$ in the sequence $x^k$. Then, the following holds
    \begin{equation}
        \begin{split}
            E_R (\rho^{\otimes k}) &\leq \sum_{x^k} p_{x^k} E_R(\rho_{x_1} \otimes \cdots \otimes \rho_{x_k})\\
            &= \sum_{x^k} p_{x^k} E_R(\bigotimes_x \rho_x^{\otimes N_x (x^k)})\\
            &\leq \sum_{x^k} p_{x^k} \sum_x E_R (\rho_x^{\otimes N_x (x^k)})\\
            &\leq \sum_{x^k} p_{x^k} \sum_x [N_x(x^k) (e_x + \delta) + C_x]\\
            &= k \sum_x p_x (e_x + \delta) + \sum_x C_x,
        \end{split}
    \end{equation}
    where the first inequality follows from the convexity of the REE (Eq.~(\ref{eq:REEConvexity})), the second inequality follows from the subadditivity of the REE (Eq.~(\ref{eq:REESubadditivity})), and we use that $\mathbb{E} [N_x (x^k)] = k p_x$ for the last equality. Consequently, at $\delta \rightarrow 0$,
    \begin{equation}
        E_R^\infty(\sum_x p_x \rho_x) = \lim_{k \rightarrow \infty} \frac{1}{k} E_R(\rho^{\otimes k}) \leq \sum_x p_x E_R^\infty (\rho_x).
    \end{equation}
\end{proof}

In particular, the bipartite REE has the following property:
\begin{lem}[Values of the bipartite REE on pure states]\label{lem:BipartiteREEonPureStates}
    For a pure state $\ket{\varphi}_{XY} = \sum_j \sqrt{\lambda_j} \ket{j}_X \ket{j}_Y$, we have the following relationship for the bipartite REE and the bipartite regularized REE
\begin{equation}\label{eq:BipartiteREEonPureState}
    E_R(\varphi) = E_R^\infty (\varphi) = S(X)_\varphi = S(Y)_\varphi.
\end{equation}
Moreover, for a quantum state $\rho$, we have
\begin{equation}\label{eq:BipartiteRegularizedREEUpperbound}
    0 \leq E_R^\infty (X:Y)_\rho \leq \min \{ S(X)_\rho, S(Y)_\rho  \}.
\end{equation}
\end{lem}

\begin{proof}[Proof of Lemma~\ref{lem:BipartiteREEonPureStates}]
    For a specific bipartite separable state $\sigma_* = \sum_j \lambda_j \ketbra{jj}{jj}$, the following holds
    \begin{equation}
        \begin{split}
            E_R  (\varphi) &\leq D(\varphi || \sigma_*)\\
            &= - \text{Tr}[\varphi \log_2 \sigma_*]\\
            &= - \sum_j \lambda_j \log_2 \lambda_j\\
            &= S(X)_\varphi = S(Y)_\varphi,
        \end{split}
    \end{equation}
    where the first equality holds since $\varphi$ is pure. On the other hand, for arbitrary separable state $\sigma_{XY} = \sum_k p_k \sigma_X^{(k)} \otimes \sigma_Y^{(k)}$, we have $\sigma_X^{(k)} \leq I_X$, and therefore $\sigma_{XY} \leq I_X \otimes \sigma_Y$. Then, the operator monotonicity of the logarithm gives
    \begin{equation}
        \begin{split}
            D(\rho_{XY} || \sigma_{XY}) &= \text{Tr}[\rho_{XY} \log_2 \rho_{XY}] - \text{Tr}[\rho_{XY} \log_2 \sigma_{XY}]\\
            &\geq - S(XY)_\rho - \text{Tr}[\rho_Y \log_2 \sigma_Y]\\
            &= S(Y)_\rho - S(XY)_\rho + D(\rho_Y || \sigma_Y)\\
            &\geq S(Y)_\rho - S(XY)_\rho,
        \end{split}
    \end{equation}
    where the last inequality follows from the nonnegativity of the quantum relative entropy. Since the von Neumann entropy vanishes for pure states, we have
    \begin{equation}
        E_R(\varphi)  = S(X)_\varphi = S(Y)_\varphi.
    \end{equation}
    In addition, due to the additivity of the von Neumann entropy
    \begin{equation}
        E_R^\infty (\varphi) = \lim_{n \rightarrow \infty} \frac{1}{n} E_R(\varphi^{\otimes n}) = \lim_{n \rightarrow \infty} \frac{1}{n} S(X^{\otimes n})_{\varphi^{\otimes n}} = S(X)_\varphi = S(Y)_\varphi.
    \end{equation}

    Moreover, for arbitrary pure-state decomposition of a mixed state $\rho_{XY} = \sum_i p_i \varphi_i$, we have
    \begin{equation}
        \begin{split}
            E_R (\rho_{XY}) &\leq \sum_i p_i E_R(\varphi_i)= \sum_i p_i S(X)_{\varphi_i}\leq S(X)_\rho,
        \end{split}
    \end{equation}
    where the first inequality follows from the convexity of the REE (Eq.~\eqref{eq:REEConvexity}) and the last inequality follows from the concavity of the von Neumann entropy. Similarly, $E_R (\rho_{XY}) \leq S(Y)_{\rho}$. Using the subadditivity of the REE (Eq.~\eqref{eq:REESubadditivity}) implies $E_R^\infty (\rho) \leq E_R$. Hence
    \begin{equation}
        0 \leq E_R^\infty(\rho_{XY}) \leq \min \{ S(X)_\rho, S(Y)_\rho \}.
    \end{equation}
\end{proof}

Define the conditional mutual information (CMI) of a quantum state $\rho$ in a composite system $XYZ$ as $I(X:Z|Y)_\rho \coloneqq S(XY)_\rho + S(YZ)_\rho - S(Y)_\rho - S(XYZ)_\rho$. Although several multipartite generalisations of the CMI exist~\cite{Yang2009Squashed}, here we use the standard CMI, where each of the systems $X$, $Y$, and $Z$ may consist of multiple subsystems. Then, we have the following relationship between the regularized REE and the CMI:
\begin{lem}[CMI inequality]\label{lem:CMIinequality}
    For an arbitrary $N+1$-partite quantum state $\rho_{X_1\cdots X_{N-1}YZ}$,
    \begin{equation}\label{eq:CMIinequality}
        E_R^\infty (X_1 : \cdots : X_{N-1} : YZ)_\rho - E_R^\infty(X_1: \cdots :X_{N-1}:Y)_\rho \leq I(X_1 \cdots X_{N-1}:Z|Y)_\rho.
    \end{equation}
\end{lem}
The three-partite case of the inequality was proved in Refs.~\cite{Brandao2011Faithful, Berta2024Entanglement}. We extend the proof in Ref.~\cite{Berta2024Entanglement} to a larger number of parties as follows.

\begin{proof}[Proof of Lemma~\ref{lem:CMIinequality}.]
    Let $X = X_1 \cdots X_{N-1}$. We first assume that $\rho_{XYZ}$ is full rank. Let $\sigma_{XY}^*$ denote an optimal fully-separable state for $E_R(X_1: \cdots : X_{N-1}:Y)$, expressed as $\sigma_{XY}^* \coloneqq \sum_a p_a \sigma_{X_1}^a \otimes \cdots \otimes \sigma_{X_{N-1}}^a \otimes \sigma_Y^a$. Since $E_R(X_1 : \cdots: X_{N-1}: Y) = -S(XY) - \text{Tr}[\rho_{XY} \log_2 \sigma_{XY}^*]$, the following holds
    \begin{equation}
        \begin{split}
            I(X:Z|Y)_\rho + E_R(X_1 : \cdots : X_{N-1} : Y)_\rho &= S(YZ)_\rho - S(Y)_\rho - S(XYZ)_\rho - \text{Tr}[\rho_{XY} \log_2 \sigma_{XY}^*] \\
            &= \text{Tr}[\rho_{XYZ} (\log_2 \rho_{XYZ} - \log_2 \rho_{YZ} + \log_2 \rho_Y - \log_2 \sigma_{XY}^*)]\\
            &= D(\rho_{XYZ} || \exp(\ln \rho_{YZ} + \ln \sigma_{XY}^* - \ln \rho_Y)).
        \end{split}
    \end{equation}
    Note that, although $\exp(\ln \rho_{YZ} + \ln \sigma_{XY}^* - \ln \rho_Y)$ is not necessarily a normalized density operator, for a quantum state $\rho$ and a positive operator $Q$, we can define $D(\rho||Q) \coloneqq \text{Tr}[\rho(\log_2 \rho - \log_2 Q)]$.
    Using Proposition 6 of Ref.~\cite{ Berta2024Entanglement} gives
    \begin{equation}
        D(\rho_{XYZ} || \exp(\ln \rho_{YZ} + \ln \sigma_{XY}^* - \ln \rho_Y)) \geq D_\text{ALL}(\rho_{XYZ} || (\text{id}_X \otimes \mathcal{R}_{Y \rightarrow YZ}) (\sigma_{XY}^*)),
    \end{equation}
    where $D_\text{ALL}$ is measured relative entropy defined as $D_\text{ALL} \coloneqq \sup_M D(M(\rho) || M(\sigma))$ for arbitrary POVM measurement $M$ and quantum states $\rho$ and $\sigma$, and $\mathcal{R}_{Y \rightarrow YZ}$ is a recovery channel applied to the system $Y$. Define $a_t \coloneqq \frac{1+it}{2}$, $\bar{a}_t \coloneqq \frac{1-it}{2}$, and $\mathcal{R}_t (\sigma) \coloneqq \rho_{YZ}^{a_t} \rho_{Y}^{-a_t} \sigma \rho_Y^{-\bar{a}_t} \rho_{YZ}^{\bar{a}_t}$. Then, the recovery channel is defined as $\mathcal{R}_{Y\to YZ} (\sigma) \coloneqq \int_\mathbb{R} \frac{\pi}{2(\cosh (\pi t) + 1)} \mathcal{R}_t (\sigma) \dd t$. Since $\mathcal{R}_{Y \rightarrow YZ}$ applies to only $Y$,
    \begin{equation}
        (\text{id}_X \otimes \mathcal{R}_{Y \rightarrow YZ}) (\sigma_{XY}^*) \in \text{SEP} (X_1: \cdots : X_{N-1}: YZ).
    \end{equation}
    Hence
    \begin{equation}
        D_\text{ALL} (\rho_{XYZ} || (\text{id}_X \otimes \mathcal{R}_{Y\rightarrow YZ}) (\sigma_{XY}^*)) \geq E_{R, \text{ALL}} (X_1: \cdots : X_{N-1} : YZ)_\rho,
    \end{equation}
    where $E_{R, \text{ALL}}$ is measured REE, which is defined as $E_{R, \text{ALL}} (\rho) \coloneqq \inf_\sigma D_\text{ALL} (\rho || \sigma)$ for a quantum state $\rho$ and a fully-separable state $\sigma$. Consequently, we have
    \begin{equation}
        \begin{split}
            I(X:Z|Y)_\rho + E_R(X_1: \cdots : X_{N-1} : Y)_\rho &\geq E_{R, \text{ALL}} (X_1 : \cdots : X_{N-1} : YZ)_\rho.
        \end{split}
    \end{equation}
    We apply the above inequality to $\rho_{XYZ}^{\otimes n}$. Since the CMI is additive on tensor products, $I(X^n : Z^n | Y^n)_{\rho^{\otimes n}} = n I(X:Z|Y)_\rho$. Then, we have
    \begin{equation}
        \begin{split}
            I(X:Z|Y)_\rho + \frac{1}{n} E_R(X_1^n: \cdots : X_{N-1}^n : Y^n)_{\rho^{\otimes n}} &\geq \frac{1}{n} E_{R, \text{ALL}} (X_1^n : \cdots : X_{N-1}^n : Y^nZ^n)_{\rho^{\otimes n}}.
        \end{split}
    \end{equation}
    Here, we use Lemma 1 of Ref.~\cite{Berta2024Entanglement}, i.e., for any $n$-partite quantum state $\rho_n$ and $\sigma_n \geq 0$ with $\rho_n \ll \sigma_n$, we have
    \begin{equation}
        D(\rho_n || \sigma_n) - \log_2 |\text{spec} (\sigma_n)| \leq D_\text{ALL} (\rho_n||\sigma_n) \leq D(\rho_n || \sigma_n),
    \end{equation}
    where $|\text{spec}(\sigma_n)|$ denotes the number of different eigenvalues of $\sigma_n$. In particular, when $\sigma_n$ is invariant under permutations of the $n$ systems, we have $|\text{spec} (\sigma_n)| \leq \text{poly}(n)$, i.e., $\log_2 |\text{spec} (\sigma_n)| \leq O(\log_2 n)$. For $n$ copies of $\rho$, the optimization over fully-separable states in the definition of the measured REE can be restricted to permutation-invariant states. Indeed, for an arbitrary fully-separable state $\sigma_n$, define its symmetrization as
    \begin{equation}
        \bar{\sigma}_n \coloneqq \frac{1}{n!} \sum_{\pi \in S_n} U_\pi \sigma_n U_\pi^\dagger,
    \end{equation}
    where $S_n$ is the symmetric group and $U_\pi$ represents an operation which simultaneously permutes the $n$ copies of all parties' systems. Since each $U_\pi$ is a product of local unitaries, $\bar{\sigma}_n$ is also a fully-separable state. Furthermore, by the convexity and unitary invariance of the measured relative entropy $D_\text{ALL}$, together with the permutation invariance of $\rho^{\otimes n}$, we have
    \begin{equation}
        D_\text{ALL} (\rho^{\otimes n} || \bar{\sigma}_n) \leq \frac{1}{n!} \sum_{\pi \in S_n} D_\text{ALL} (\rho^{\otimes n}||U_\pi \sigma_n U_\pi^\dagger) = \frac{1}{n!} \sum_{\pi \in S_n} D_\text{ALL} (U_\pi^\dagger \rho^{\otimes n} U_\pi||\sigma_n ) = D_\text{ALL} (\rho^{\otimes n} || \sigma_n).
    \end{equation}
    Therefore, the optimization of the measured REE can be restricted to permutation-invariant fully-separable states without changing its value. For such permutation-invariant states, we have $|\text{spec} (\bar{\sigma}_n)|\leq \text{poly} (n)$. Hence, Lemma 1 of Ref.~\cite{Berta2024Entanglement} implies
    \begin{equation}
        E_{R, \text{ALL}} (\rho^{\otimes n}) \geq E_R(\rho^{\otimes n}) - O(\log_2 n).
    \end{equation}
    Using $D_\text{ALL} (\rho_n || \sigma_n) \leq D(\rho_n||\sigma_n)$, gives $E_{R, \text{ALL}}(\rho^{\otimes n}) \leq E_R(\rho^{\otimes n})$. Then, dividing by $n$ and taking $n \to \infty$ gives
    \begin{equation}
        \lim_{n \to \infty} \frac{1}{n} E_{R, \text{ALL}} (\rho^{\otimes n}) = \lim_{n \rightarrow \infty} \frac{1}{n} E_R (\rho^{\otimes n}).
    \end{equation}
    Hence, we have
    \begin{equation}
        \begin{split}
            I(X:Z|Y)_\rho + E_R^\infty(X_1: \cdots : X_{N-1} : Y)_\rho &\geq E_R^\infty (X_1 : \cdots : X_{N-1} : YZ)_\rho.
        \end{split}
    \end{equation}
    That is
    \begin{equation}
        E_R^\infty (X_1 : \cdots : X_{N-1}: YZ)_\rho - E_R^\infty (X_1 : \cdots : X_{N-1} : Y)_\rho \leq I(X_1 \cdots X_{N-1} : Z|Y)_\rho.
    \end{equation}

    For a $d$-dimensional quantum state $\rho$ that does not have full support, we consider $\rho_\delta = (1- \delta) \rho + \delta I/d$. By taking $\delta \rightarrow 0$, due to the continuity of $E_R^\infty$ which we describe later, the same inequality holds.
\end{proof}

Finally we describe the continuity of the REE and the regularized REE. To this end, we introduce the following functions:
\begin{equation}\label{eq:EntropyFunctions}
    \begin{split}
        h(p) &\coloneqq -p \log_2 p - (1-p) \log_2 (1-p),\\
        g(p) &\coloneqq (1+p)h(\frac{p}{1+p})= (1+p)\log_2 (1+p) - p\log_2 p,\\
        b_N(p) &\coloneqq h(p) + \frac{1}{N-1} g(p),
    \end{split}
\end{equation}
where $h$ is the binary entropy function and $b_N$ is defined for $N \geq 2$. Then, we have the following lemma:

\begin{lem}[Continuity of the regularized REE]\label{lem:RegularizedREEContinuity}
    For two quantum states $\rho$ and $\sigma$ on a Hilbert space of dimension $D$, satisfying $T(\rho, \sigma) \leq t \leq 1$, we have
    \begin{equation}\label{eq:VonNeumannContinuity}
        |S(\rho) - S(\sigma)| \leq t \log_2 D + h(t).
    \end{equation}
    Furthermore, for $N$-partite states whose local dimensions are at most $d$, we have
    \begin{equation}\label{eq:REEContinuity}
        |E_R (\rho) - E_R (\sigma)| \leq t (N-1) \log_2 d + g(t),
    \end{equation}
    and
    \begin{equation}\label{eq:RegularizedREEContinuity}
        |E_R^\infty (\rho) - E_R^\infty (\sigma)| \leq t (N-1) \log_2 d + g(t).
    \end{equation}
\end{lem}

\begin{proof}[Proof of Lemma~\ref{lem:RegularizedREEContinuity}]
    See, e.g., Ref.~\cite{Winter2016Tight} for the proof of the continuity of the von Neumann entropy. Consider an $N$-partite state $\rho_{A_1 \cdots A_N}$ with local dimension at most $d$. Let $\rho'_{A_1\cdots A_N}$ denote a fully-separable state after we dephase $N-1$ systems of $\rho$. Then,
    \begin{equation}
        \begin{split}
            D(\rho || \rho') &= -\text{Tr}[\rho \log_2 \rho'] - S(\rho)= S(\rho') - S(\rho),
        \end{split}
    \end{equation}
    where the last equality holds since $\log_2 \rho'$ is a block-diagonal operator kept under the dephasing. Dephasing a $d$-dimensional system increases the von Neumann entropy at most $\log_2 d$. Therefore, we have $S(\rho') - S(\rho) \leq (N-1) \log_2 d$. Since $\rho'$ is a fully-separable state, $E_R(\rho) \leq D(\rho || \rho')\leq (N-1) \log_2 d$. Consequently, due to the nonnegativity of the REE (Eq.~\eqref{eq:REENonnegativity}), we have
    \begin{equation}\label{eq:E_RRange}
        0 \leq E_R(\rho) \leq (N-1) \log_2 d.
    \end{equation}

    For a convex set $\mathcal{C}$, define $D_\mathcal{C} (\rho) \coloneqq \inf_{\tau \in \mathcal{C}} D(\rho || \tau)$. Let $\kappa$ denote the difference between the maximum and the minimum of $D_\mathcal{C}$, i.e., $\kappa \coloneqq \sup_{\rho, \sigma} [D_\mathcal{C} (\rho) - D_\mathcal{C} (\sigma)]$. Using Lemma 7 in Ref.~\cite{Winter2016Tight}, when $\mathcal{C}$ is a closed, convex and bounded set of positive semidefinite operators, containing at least one full-rank operator, we have $|D_\mathcal{C} (\rho) - D_\mathcal{C} (\sigma)| \leq t \kappa + g(t)$ for any quantum states $\rho$ and $\sigma$ with $T(\rho, \sigma) \leq t$. In particular, by taking $\mathcal{C} = \text{SEP}$, $D_\mathcal{C} (\rho) = E_R(\rho)$ and $\kappa \leq (N-1) \log_2 d$ because of Eq.~\eqref{eq:E_RRange}. Consequently,
    \begin{equation}
        |E_R(\rho) - E_R(\sigma)| \leq t (N-1) \log_2 d + g(t).
    \end{equation}

    Finally, we prove the continuity of the regularized REE. To this end, we generalize the proof of Corollary 8 in Ref.~\cite{Winter2016Tight}. We first consider arbitrary $N$-partite states $\tau$ and $\Omega$. Here, the local dimension of $\tau$ is at most $d$. Due to the LOCC monotonicity of the REE (Eq.~\eqref{eq:REELOCCmonotone}), we have $E_R(\tau \otimes \Omega) \geq E_R(\Omega)$. On the other hand, since the REE is subadditive (Eq.~\eqref{eq:REESubadditivity}),
    \begin{equation}\label{eq:ERtauOmega}
        \begin{split}
            E_R(\tau \otimes \Omega) &\leq E_R(\tau) + E_R(\Omega)\leq (N-1) \log_2 d + E_R(\Omega).
        \end{split}
    \end{equation}
    Applying the argument in the proof of Corollary 8 in Ref.~\cite{Winter2016Tight} to $E_R(\tau \otimes \Omega)$, whose range has width at most $(N-1)\log_2 d$ by Eq.~\eqref{eq:ERtauOmega}, gives
    \begin{equation}
        |E_R (\rho \otimes \Omega) - E_R (\sigma \otimes \Omega)| \leq t (N-1) \log_2 d + g(t).
    \end{equation}
    Define $\Omega_j \coloneqq \rho^{\otimes (j-1)} \otimes \sigma^{\otimes (n -j)}$. Then,
    \begin{equation}
        \begin{split}
            |E_R(\rho^{\otimes n}) - E_R(\sigma^{\otimes n})| &= |\sum_{j = 1}^n [  E_R(\rho \otimes \Omega_j) - E_R(\sigma \otimes \Omega_j)]|\\
            &\leq \sum_{j = 1}^n |E_R (\rho \otimes \Omega_j) - E_R (\sigma \otimes \Omega_j)|\\
            &\leq n [t (N-1) \log_2 d + g(t)],
        \end{split}
    \end{equation}
    where we use the triangle inequality for the first inequality. Consequently,
    \begin{equation}
        |E_R^\infty (\rho) - E_R^\infty(\sigma)| \leq t (N-1) \log_2 d + g(t).
    \end{equation}
\end{proof}

\section{B. Marginal Entanglement Balance (MEB)}
\subsection{1. Definition and basic properties}
In this section, we introduce the marginal entanglement balance (MEB). We first focus on pure states and describe an extension to mixed states later. The definition of the MEB for pure states is as follows:

\begin{dfn}[Marginal entanglement balance on pure states]
    Consider an $N\geq2$-partite pure state $\psi_{A_1 \cdots A_N}$. Let $\mathcal{P} = \{ P_1, \ldots, P_m \}$ be a partition of the $N$ parties into $m = |\mathcal{P}| \geq 2$ blocks. Then, marginal entanglement balance (MEB) on pure states $M_{N, \infty}$ is defined as follows
    \begin{equation}\label{eq:MEBDefinition}
        M_{N, \infty} (\psi) \coloneqq \min_{\mathcal{P} : |\mathcal{P}|\geq 2} M_{\mathcal{P}, \infty}(\psi),
    \end{equation}
    where the minimization is over all partitioning $\mathcal{P}$ and
    \begin{equation}\label{eq:MEBDefinitionFixed}
        M_{\mathcal{P}, \infty} (\psi) \coloneqq \frac{1}{m} \sum_{i = 1}^m S(P_i)_\psi - \frac{1}{m(m-1)} \sum_{i = 1}^{m} E_{\bar{i}} (\psi).
    \end{equation}
    Here, $S(P_i)_\psi$ represents the marginal von Neumann entropy of block $P_i$, $\bar{i} = [m]\backslash\{ i\}$ and 
    \begin{equation}\label{eq:EBari}
        E_{\bar{i}} (\psi) \coloneqq E_R^\infty (P_1 : \cdots : \hat{P_i} : \cdots P_m)_{\psi_{\bar{i}}},
    \end{equation}
    where $\hat{P_i}$ means that we exclude the block $P_i$ and $\psi_{\bar{i}}$ denotes the reduced state obtained by tracing out the block $P_i$.
\end{dfn}

The MEB for pure states $M_{N, \infty}$ has the following basic properties:

\begin{lem}[Basic properties of MEB]\label{lem:MEBBasicProperties}
For an $N$-partite pure state $\psi_{A_1 \cdots A_{N}}$ and fixed partitioning $\mathcal{P}=\{P_1, \ldots, P_m \}$,
\begin{equation}\label{eq:MEBNonnegativityFixed}
    0 \leq \frac{1}{m(m-1)} \sum_i S(P_i)_\psi \leq M_{\mathcal{P}, \infty}(\psi) \leq \frac{1}{m}\sum_i S(P_i)_\psi,
\end{equation}
and
\begin{equation}\label{eq:MEBNonnegativity}
    0 \leq \min_{\mathcal{P}:|\mathcal{P}|\geq 2}\frac{1}{m(m-1)} \sum_i S(P_i)_\psi \leq M_{N, \infty}(\psi) \leq \min_{\mathcal{P}:|\mathcal{P}|\geq 2}\frac{1}{m}\sum_i S(P_i)_\psi.
\end{equation}

In particular, when $\psi$ is a biseparable pure state, $M_{N, \infty}(\psi) = 0$. In addition, for the $d$-dimensional $N$-partite GHZ state $\Gamma_d^{(N)}$,
\begin{equation}
    M_{N, \infty} (\Gamma_d^{(N)}) = \log_2 d.
\end{equation}

The MEB is invariant under local isometries and it is weakly additive on tensor products, i.e.,
\begin{equation}
    M_{N, \infty} (\psi^{\otimes k} ) = k M_{N, \infty} (\psi).
\end{equation}
\end{lem}

\begin{proof}[Proof of Lemma~\ref{lem:MEBBasicProperties}]
    For an $m$-partite pure state $\psi_{P_1 \cdots P_{m}}$,
    \begin{equation}
        \begin{split}
            E_{\bar{i}} (\psi) 
            &\leq E_R(\psi_{\bar{i}})\\
            &\leq D(\psi_{\bar{i}} || \bigotimes_{j \neq i} \psi_{P_j})\\
            &= - S(\psi_{\bar{i}}) - \sum_{j \neq i} \text{Tr} [\psi_{P_j} \log_2 \psi_{P_j}]\\
            &= \sum_{j \neq i} S(P_j)_\psi - S(P_i)_\psi\\
            &= \sum_j S(P_j) - 2 S(P_i),
        \end{split}
    \end{equation}
    where the first inequality follows from the definition of the regularized REE and the subadditivity of the REE (Eq.~(\ref{eq:REESubadditivity})), the second inequality follows from the definition of the REE, and the third equality holds because $\psi$ is a pure state and $S(\psi_{\bar{i}}) = S(\psi_{P_i})$. Consequently,
    \begin{equation}
        \begin{split}
            \sum_{i = 1}^m E_{\bar{i}} (\psi) &\leq m \sum_j S(P_j) - 2 \sum_i S(P_i)= (m-2) \sum_{i = 1}^m S(P_i).
        \end{split}
    \end{equation}
    Therefore, we have
    \begin{equation}
        \begin{split}
            M_{\mathcal{P}, \infty} (\psi) &= \frac{1}{m} \sum_i S(P_i) - \frac{1}{m(m-1)} \sum_i E_{\bar{i}} (\psi)\\
            &\geq \frac{1}{m(m-1)} \sum_{i} S(P_i)\\
            &\geq 0,
        \end{split}
    \end{equation}
    where the last inequality holds since the von Neumann entropy is nonnegative. On the other hand, since the REE is nonnegative (Eq.~(\ref{eq:REENonnegativity})),
    \begin{equation}
        M_{\mathcal{P}, \infty} (\psi) \leq \frac{1}{m} \sum_i S(P_i).
    \end{equation}
    Then, Eq.~\eqref{eq:MEBNonnegativity} is obtained from the definition of the MEB in Eq.~\eqref{eq:MEBDefinition}.
    
    When $\psi$ is a biseparable pure state, for the corresponding bipartition, the first term of the MEB becomes zero since the von Neumann entropy vanishes for pure states. In addition, since $\psi$ is biseparable, the second term of the MEB also vanishes. Consequently, $M_{N, \infty} (\psi) = 0$. For the GHZ state $\Gamma_d^{(N)}$, for arbitrary partitioning $\mathcal{P} = \{  P_1, \ldots, P_m\}$, the von Neumann entropy of each block $P_i$ becomes $S(P_i) = \log_2 d$. When we trace out one block from $\Gamma_d^{(N)}$, the reduced state is fully separable. Therefore, $M_{N, \infty}(\Gamma_d^{(N)}) = \log_2 d$.

    Since both the von Neumann entropy and the regularized REE are invariant under local isometries, the MEB is also invariant under local isometries. Using the weak additivity of the von Neumann entropy and the regularized REE (Eq.~\eqref{eq:RegularizedREEWeakAdditivity}), we conclude that the MEB is also weakly additive, i.e.,
    \begin{equation}
        \begin{split}
            M_{N, \infty}(\psi^{\otimes k}) = \min_{\mathcal{P}:|\mathcal{P}|\geq 2} M_{\mathcal{P}, \infty}(\psi^{\otimes k}) = k \min_{\mathcal{P}:|\mathcal{P}|\geq 2}M_{\mathcal{P}, \infty} (\psi) = kM_{N,\infty}(\psi).
        \end{split}
    \end{equation}
\end{proof}

\subsection{2. LOCC monotonicity}
In this subsection, we show that the MEB on pure states is LOCC monotone:

\begin{lem}[LOCC monotonicity of MEB on pure states]\label{lem:PureMEBLOCCMonotone}
    Consider a local quantum instrument performed by one of the $N$ parties, characterized by $\{ p_x, \psi^x \}$, where $p_x$ is the probability of obtaining outcome $x$ and $\psi^x$ is a pure quantum state after the measurement. Then, the MEB on pure states is LOCC monotone:
    \begin{equation}\label{eq:PureMEBLOCCMonotone}
        M_{N, \infty} (\psi) \geq \sum_x p_x M_{N, \infty} (\psi^x).
    \end{equation}
\end{lem}

\begin{proof}[Proof of Lemma~\ref{lem:PureMEBLOCCMonotone}]
    Consider a fixed partition $\mathcal{P} = \{  P_1, \ldots, P_m\}$. Suppose block $P_k$ performs a local quantum instrument. Let $P_k'$ denote the system of the block after the measurement. Define $\delta_i \coloneqq S(P_i)_\psi - \sum_x p_x S(P_i)_{\psi^x}$ where, for $i = k$, the second term becomes $\sum_x p_xS(P_k')_{\psi^x}$. Due to the concavity of the von Neumann entropy, for $i \neq k$,
    \begin{equation}\label{eq:delta_iNonnegativity}
        \delta_i \geq 0 \quad (i \neq k).
    \end{equation}
    In addition, define
    \begin{equation}\label{Eq:L_iDefinition}
        L_i \coloneqq E_{\bar{i}} (\psi) - \sum_x p_x E_{\bar{i}} (\psi^x).
    \end{equation}
    When $i = k$, the reduced state of the unmeasured parties is unchanged after averaging over the measurement outcomes. Therefore, due to the convexity of the regularized REE (Eq.~\eqref{eq:RegularizedREEConvexity}), we have $L_k \leq 0$.

    In the following, we derive upper bounds on $L_i$ for $i \neq k$. We first consider $P_k$'s local isometric dilation involving two auxiliary registers $X$ and $X'$: $\ket{\Omega}_{P_{\bar{k}} P_k' XX'}
        =\sum_x \sqrt{p_x} \ket{\psi^x}_{P_{\bar{k}} P_k'} \ket{x}_X \ket{x}_{X'}$. Tracing out the register $X'$ gives $\Omega_{P_{\bar{k}} P_k'X} = \sum_x p_x \ketbra{x}{x}_X \otimes \psi^x_{P_{\bar{k}}P_k'}$. Since the regularized REE is invariant under local isometries, we have $E_{\bar{i}} (\{ P_j \}_{j \neq i})_\psi = E_{\bar{i}} (\{P_j  \}_{j \neq i, k}:P_k'XX')_\Omega$. Using the convexity of the regularized REE (Eq.~(\ref{eq:RegularizedREEConvexity})) gives
    \begin{equation}\label{eq:AXBConvexity}
        \begin{split}
            E_{\bar{i}} (\{P_j  \}_{j \neq i, k} : P_k' X)_\Omega &\leq \sum_x p_x E_{\bar{i}} (\psi^x_{P_{\bar{k}}P_k'} \otimes \ketbra{x}{x}_X)=\sum_x p_x E_{\bar{i}} (\psi^x),
        \end{split}
    \end{equation}
    where the last equality holds because the regularized REE is invariant under adding local ancillas. Therefore, we have the following
    \begin{equation}\label{eq:L_iUpperCMI}
        \begin{split}
            L_{i} &= E_{\bar{i}} (\psi) - \sum_x p_x E_{\bar{i}} (\psi^x)\\
            &\leq E_{\bar{i}} (\{P_j  \}_{j \neq i, k}:P_k'XX')_\Omega - E_{\bar{i}} (\{P_j  \}_{j \neq i, k} : P_k' X)_\Omega\\
            &= E_R ^\infty(\{P_j  \}_{j \neq i, k}:P_k'XX')_\Omega - E_R^\infty (\{P_j  \}_{j \neq i, k} : P_k' X)_\Omega\\
            &\leq I(P_{[m]\backslash \{i, k  \}  }:X'|P_k'X)_\Omega,
        \end{split}
    \end{equation}
    where the first equality follows from the definition of $L_{i}$ (Eq.~(\ref{Eq:L_iDefinition})), the first inequality follows from Eq.~(\ref{eq:AXBConvexity}), the last equality follows from Eq.~\eqref{eq:EBari} and the last inequality follows from the CMI inequality in Eq.~(\ref{eq:CMIinequality}). For $m=2$, $P_{[m]\backslash \{i,k \}}$ becomes an empty set and we take $I(P_{[m] \backslash \{i, k \}}: X'|P_k' X)_\Omega = 0$. Since $\Omega_{P_{\bar{k}}P_k'X}$ is a classical-quantum block diagonal state and $\ket{\Omega}_{P_{\bar{k}} P_k' XX'}$ and $\psi$ are pure states, we have
    \begin{equation}
        \begin{split}
            S(P_{[m]\backslash \{ i, k \}}P_k'X)_\Omega &= S(P_i X')_\Omega = H(p_x) + \sum_x p_x S(P_i)_{\psi^x},\\
            S(P_k'X)_\Omega &= H(p_x) + \sum_x p_x S(P'_k)_{\psi^x},\\
            S(P_k'XX')_\Omega &= S(P_{[m]\backslash \{ k\}})_\Omega = S(P_{[m] \backslash \{k\}})_\psi = S(P_k)_\psi,\\
            S(P_{[m]\backslash \{ i, k \}}P'_kXX')_\Omega &= S(P_i)_\Omega = S(P_i)_\psi.
        \end{split}
    \end{equation}
    where $H$ is the Shannon entropy. Hence, the CMI becomes
    \begin{equation}
        \begin{split}
            I(P_{[m]\backslash \{ i, k \}}:X'|P_k'X)_\Omega &= S(P_{[m]\backslash \{ i, k \}}P_k'X)_\Omega + S(P_k'XX')_\Omega - S(P_k'X)_\Omega - S(P_{[m]\backslash \{ i, k \}}P_k'XX')_\Omega\\
            &= \sum_x p_x S(P_i)_{\psi^x} + S(P_k)_\psi - \sum_x p_x S(P_k')_{\psi^x} - S(P_i)_\psi\\
            &= \delta_k - \delta_i.
        \end{split}
    \end{equation}
    Consequently,
    \begin{equation}\label{eq:LtotalUpperbound}
    \begin{split}
        \sum_i L_i \leq \sum_{i \neq k} L_i \leq (m-1) \delta_k - \sum_{i \neq k} \delta_i
    \end{split}
    \end{equation}
    Therefore, from the definition of the MEB (Eq.~(\ref{eq:MEBDefinitionFixed})), we have
    \begin{equation}
        \begin{split}
            M_{\mathcal{P}, \infty}(\psi) - \sum_x p_x M_{\mathcal{P}, \infty} (\psi^x) &= \frac{1}{m}\sum_{i = 1}^m S(P_i)_\psi - \frac{1}{m(m-1)}\sum_{i = 1}^m E_{\bar{i}} (\psi)\\
            &-\sum_x p_x \{\frac{1}{m}\sum_{i = 1}^m S(P_i)_{\psi^x} - \frac{1}{m(m-1)}\sum_{i = 1}^m E_{\bar{i}} (\psi^x) \}\\
            &= \frac{1}{m} \sum_{i = 1}^m \delta_i  - \frac{1}{m(m-1)} \sum_{i = 1}^m L_i\\
            &\geq \frac{1}{m} \sum_{i = 1}^m \delta_i - \frac{1}{m(m-1)} ((m-1)\delta_k - \sum_{ i \neq k} \delta_i)\\
            &= \frac{1}{m-1} \sum_{ i \neq k} \delta_i\\
            &\geq 0,
        \end{split}
    \end{equation}
    where the last inequality follows from Eq.~(\ref{eq:delta_iNonnegativity}). Since any LOCC protocol among the original parties is also an LOCC protocol with respect to an arbitrary partition, the above inequality holds for any fixed partition $\mathcal{P}$. Let $\mathcal{P}^*$ be a partition minimizing the MEB. Then,
    \begin{equation}
        \begin{split}
            M_{N,\infty} (\psi) &= M_{\mathcal{P}^*, \infty} (\psi) \geq \sum_x p_x M_{\mathcal{P}^*, \infty} (\psi^x) \geq \sum_x p_x \min_{\mathcal{P}: |\mathcal{P}|\geq 2} M_{\mathcal{P}, \infty} (\psi^x) = \sum_x p_x M_{N, \infty} (\psi^x).
        \end{split}
    \end{equation}
    This concludes the proof of the LOCC monotonicity of the MEB on pure states.
\end{proof}

\subsection{3. Convex-roof extension}
So far, $M_{\mathcal{P}, \infty}$ and $M_{N, \infty}$ have been defined for pure states. In this subsection, we extend their definitions to mixed states. For a mixed $N$-partite quantum state $\rho$, define
\begin{equation}\label{eq:MEBhatDefinitionFixed}
    \hat{M}_{\mathcal{P}, \infty} (\rho) \coloneqq \inf_{\rho = \sum_i q_i \psi_i} \sum_i q_i M_{\mathcal{P}, \infty} (\psi_i),
\end{equation}
and
\begin{equation}\label{eq:MEBhatDefinition}
    \hat{M}_{N, \infty} (\rho) \coloneqq \inf_{\rho = \sum_i q_i \psi_i} \sum_i q_i M_{N, \infty} (\psi_i),
\end{equation}
where the infima are taken over all pure-state decompositions of $\rho$. Note that, for a fixed partition $\mathcal{P}'$, we have
\begin{equation}
    \begin{split}
        \hat{M}_{N,\infty}(\rho) = \inf_{\rho = \sum_i q_i \psi_i} \sum_i q_i \min_{\mathcal{P}:|\mathcal{P}|\geq 2} M_{\mathcal{P}, \infty} (\psi_i) \leq \inf_{\rho = \sum_i q_i \psi_i} \sum_i q_i M_{\mathcal{P}', \infty} (\psi_i) = \hat{M}_{\mathcal{P}', \infty} (\rho).
    \end{split}
\end{equation}
Hence
\begin{equation}\label{eq:MhatInequality}
    \hat{M}_{N, \infty} (\rho) \leq \min_{\mathcal{P} : |\mathcal{P}|\geq 2} \hat{M}_{\mathcal{P}, \infty} (\rho).
\end{equation}

In the following, we prove the basic properties and LOCC monotonicity of $\hat{M}_{N, \infty}$. The same arguments apply to $\hat{M}_{\mathcal{P}, \infty}$ for any fixed partition $\mathcal{P}$. $\hat{M}_{N, \infty}$ has the following basic properties:

\begin{lem}[Basic properties of $\hat{M}_{N, \infty}$]\label{lem:MEBhatBasicProperties}
    For arbitrary quantum state $\rho$, $\hat{M}_{N, \infty}$ is nonnegative. For arbitrary pure state $\psi$, $\hat{M}_{N, \infty}$ equals $M_{N, \infty}$, i.e., 
    \begin{equation}
        \hat{M}_{N, \infty} (\psi) = M_{N, \infty} (\psi).
    \end{equation}
    In particular, for the GHZ state,
    \begin{equation}
        \hat{M}_{N, \infty} (\Gamma_d^{(N)}) = \log_2 d.
    \end{equation}
    In addition, $\hat{M}_{N, \infty}$ is a convex function:
    \begin{equation}\label{eq:MEBhatConvexity}
        \hat{M}_{N, \infty} (\sum_x p_x \rho_x) \leq \sum_x p_x \hat{M}_{N, \infty} (\rho_x).
    \end{equation}
    Moreover, for a pure state $\psi$,
    \begin{equation}\label{eq:MEBhatWeakAdditivity}
        \hat{M}_{N, \infty} (\psi^{\otimes n}) = n M_{N, \infty} (\psi).
    \end{equation}
\end{lem}

\begin{proof}[Proof of Lemma~\ref{lem:MEBhatBasicProperties}]
    The nonnegativity of $\hat{M}_{N, \infty}$ follows from that of $M_{N, \infty}$. By the definition of $\hat{M}_{N, \infty}$, it is straightforward to see that it is convex and equals $M_{N, \infty}$ for a pure state. Since the GHZ state $\Gamma_d^{(N)}$ is a pure state and $M_{N, \infty}(\Gamma_d^{(N)}) = \log_2 d$, we have $\hat{M}_{N, \infty} (\Gamma_d^{(N)}) = \log_2 d$. In addition, since $M_{N, \infty}$ is weakly additive on tensor products, we have $\hat{M}_{N, \infty} (\psi^{\otimes n}) = M_{N, \infty} (\psi^{\otimes n}) = n M_{N, \infty}(\psi)$.
\end{proof}

We now prove that $\hat{M}_{N, \infty}$ is LOCC monotone:

\begin{prop}[LOCC monotonicity of $\hat{M}_{N, \infty}$]\label{Prop:MEBhatLOCCMonotone}
    Consider a local quantum instrument performed by one of the $N$ parties, characterized by $\{ p_x, \rho_x \}$, where $p_x$ is the probability of obtaining outcome $x$ and $\rho_x$ is a quantum state after the measurement. Then, 
    \begin{equation}\label{eq:MEBhatLOCCMonotone}
        \hat{M}_{N,\infty} (\rho) \geq \sum_x p_x \hat{M}_{N,\infty}(\rho_x).
    \end{equation}
\end{prop}

\begin{proof}[Proof of Proposition~\ref{Prop:MEBhatLOCCMonotone}]
    For arbitrary $\zeta > 0$, consider a pure-state decomposition $\rho = \sum_i q_i \psi_i$, satisfying
    \begin{equation}\label{eq:MEBhatZeta}
        \sum_i q_i M_{N,\infty} (\psi_i) \leq \hat{M}_{N,\infty} (\rho) + \zeta.
    \end{equation}
    Let $\alpha$ be a label distinguishing Kraus operators $K_{x\alpha}$ for the same outcome $x$. By refining the outcome $x$ into $(x,\alpha)$, each refined outcome is associated with a single Kraus operator. Let $\psi_{ix\alpha}$ denote a pure state obtained by performing $K_{x\alpha}$ on $\psi_i$ and let $p(x, \alpha|i)$ denote a probability distribution satisfying $p(x, \alpha|i) \coloneqq \text{Tr}[K_{x\alpha} \psi_i K_{x \alpha}^\dagger]$. Then, the unnormalized post-measurement state corresponding to outcome $x$ is given by $p_x \rho_x = \sum_{i, \alpha} q_i p(x,\alpha|i) \psi_{ix\alpha}$. Here, since $\sum_{i, \alpha} [q_i p(x, \alpha|i)]/p_x  = 1$, the following is a pure-state decomposition of $\rho_x$
    \begin{equation}
        \rho_x = \sum_{i, \alpha} \frac{q_i p(x, \alpha|i)}{p_x} \psi_{ix \alpha}.
    \end{equation}
    By the definition of $\hat{M}_{N, \infty}$, we have $p_x \hat{M}_{N, \infty} (\rho_x) \leq \sum_{i, \alpha} q_i p(x, \alpha|i) M_{N, \infty} (\psi_{ix\alpha})$. Using Lemma~\ref{lem:PureMEBLOCCMonotone} gives
    \begin{equation}
        \begin{split}
            \sum_x p_x \hat{M}_{N, \infty}(\rho_x) &\leq \sum_i q_i \sum_{x, \alpha} p(x, \alpha|i) M_{N, \infty} (\psi_{ix\alpha})\\
            &\leq \sum_i q_i M_{N, \infty} (\psi_i)\\
            &\leq \hat{M}_{N, \infty} (\rho) + \zeta,
        \end{split}
    \end{equation}
    where the second inequality follows from Lemma~\ref{lem:PureMEBLOCCMonotone} and the last inequality follows from Eq.~(\ref{eq:MEBhatZeta}). Taking $\zeta \rightarrow 0$ concludes the proof.
\end{proof}

By applying Proposition~\ref{Prop:MEBhatLOCCMonotone} successively to each round of a finite-round LOCC protocol and averaging over the measurement outcomes, we obtain the average LOCC monotonicity of $\hat{M}_{N, \infty}$. Together with the convexity of $\hat{M}_{N, \infty}$ (Eq.~\eqref{eq:MEBhatConvexity}), this implies that $\hat{M}_{N, \infty}$ is nonincreasing under finite-round LOCC channels, i.e.,
\begin{equation}\label{eq:MEBhatLOCCMonotoneLambda}
    \hat{M}_{N, \infty} (\Lambda(\rho)) = \hat{M}_{N, \infty}(\sum_x p_x \rho_x) \leq \sum_x p_x \hat{M}_{N, \infty} (\rho_x) \leq \hat{M}_{N, \infty} (\rho).
\end{equation}

\subsection{4. Continuity}
In this subsection, we describe the continuity of $\hat{M}_{N, \infty}$. We first prove the following lemma for the MEB on pure states $M_{\mathcal{P}, \infty}$ for a fixed partition $\mathcal{P}$:

\begin{lem}[Continuity of $M_{N, \infty}$]\label{lem:MEBExplicitLowerBound}
    Consider two $N$-partite pure states $\phi$ and $\tau$ satisfying $T(\phi, \tau) = t$. Then, the following holds  \begin{equation}\label{eq:MEBExplicitLowerBound}
        M_{N, \infty} (\phi) \geq M_{N, \infty}(\tau) -\tilde{\beta}_N t \log_2 d - \tilde{b}(t),
    \end{equation}
    where $d$ is an upper bound on the dimension of each local subsystem, $\tilde{\beta}_N \coloneqq 2(N-1)$, and $\tilde{b}(t) \coloneqq h(t) + g(t)$.
\end{lem}

\begin{proof}[Proof of Lemma~\ref{lem:MEBExplicitLowerBound}]
    Consider a fixed partition $\mathcal{P} = \{ P_1, \ldots, P_m \}$. Let $\phi_{P_i}$ and $\tau_{P_i}$ denote reduced states of $\phi$ and $\tau$ for block $P_i$, respectively. Due to the monotonicity of the trace distance, $T(\phi_{P_i}, \tau_{P_i}) \leq T(\phi, \tau)  = t$. Using the continuity of the von Neumann entropy in Eq.~\eqref{eq:VonNeumannContinuity} gives $|S(\phi_{P_i}) - S(\tau_{P_i})| \leq t \log_2 d_\mathcal{P} + h(t)$, where $d_\mathcal{P} \coloneqq \max_{P_i \in \mathcal{P}} \dim \mathcal{H}_{P_i}$ is the maximal dimension of each block. Then, we have $S(\phi_{P_i}) \geq S(\tau_{P_i}) -t\log_2 d_\mathcal{P} - h(t)$.

    Let $\phi_{\bar{i}}$ and $\tau_{\bar{i}}$ represent reduced states obtained by tracing out $P_i$ from $\phi$ and $\tau$, respectively. Again, due to the monotonicity of the trace distance, we have $T(\phi_{\bar{i}}, \tau_{\bar{i}}) \leq T(\phi, \tau) = t$. Using the continuity of the regularized REE (Eq.~\eqref{eq:RegularizedREEContinuity}) gives $|E_R^\infty (\phi_{\bar{i}}) - E_R^\infty(\tau_{\bar{i}})| \leq t (m-2) \log_2 d_\mathcal{P} + g(t)$. Then, we have $E_R^\infty (\phi_{\bar{i}}) \leq E_R^\infty(\tau_{\bar{i}}) + t (m-2) \log_2 d_\mathcal{P} + g(t)$. Consequently,
    \begin{equation}\label{eq:MContinuityFixed}
        \begin{split}
            M_{\mathcal{P}, \infty}(\phi) &= \frac{1}{m} \sum_i S(\phi_{P_i}) - \frac{1}{m(m-1)} \sum_i E_{R}^\infty (\phi_{\bar{i}})\\
            &\geq \frac{1}{m} \sum_i S(\tau_{P_i}) -t\log_2 d_\mathcal{P} - h(t) - \frac{1}{m(m-1)} (\sum_i E_R^\infty(\tau_{\bar{i}})+m(m-2)t\log_2 d_\mathcal{P}+m g(t))\\
            &= M_{\mathcal{P}, \infty}(\tau) -\beta_m t \log_2 d_\mathcal{P} - b_m (t),
        \end{split}
    \end{equation}
    where $\beta_m \coloneqq (2m-3)/(m-1)$ and $b_m(t) = h(t) + g(t)/(m-1)$ in Eq.~\eqref{eq:EntropyFunctions}.

    When the dimension of each subsystem of the original parties is at most $d$, the dimension of block $P_i$ is $\dim \mathcal{H}_{P_i} = \prod_{A_j \in P_i} \dim \mathcal{H}_{A_j} \leq d^{|P_i|}$. Since there are at least two blocks, we have $d_\mathcal{P} = \max_{P_i \in \mathcal{P}} \dim \mathcal{H}_{P_i} \leq d^{N-1}$. In addition, for arbitrary $m \geq 2$, we have $\beta_m = 2 -1/(m-1) < 2$ and $b_m(t) \leq h(t) + g(t)$. Substituting these into Eq.~\eqref{eq:MContinuityFixed} gives
    \begin{equation}
    \begin{split}
        M_{\mathcal{P}, \infty} (\phi) &\geq M_{\mathcal{P}, \infty} (\tau) - 2 (N-1)t \log_2 d - h(t) -g(t)\\
        &= M_{\mathcal{P}, \infty} (\tau)-\tilde{\beta}_N t \log_2 d - \tilde{b}(t).
    \end{split}
    \end{equation}
    Since this inequality holds for arbitrary partition,
    \begin{equation}
        \begin{split}
            M_{N, \infty}(\phi) = \min_{\mathcal{P}:|\mathcal{P}|\geq 2} M_{\mathcal{P}, \infty}(\phi) \geq \min_{\mathcal{P}:|\mathcal{P}|\geq 2} M_{\mathcal{P}, \infty}(\tau) - \tilde{\beta}_N t \log_2 d - \tilde{b}(t) = M_{N, \infty}(\tau) - \tilde{\beta}_Nt \log_2 d - \tilde{b}(t).
        \end{split}
    \end{equation}
    This concludes the proof.
\end{proof}

We extend the above lemma to $\hat{M}_{N, \infty}$ as follows. Note that, in the following proposition, we assume that a target state is pure. However, this is sufficient for the cases which we consider in this work.

\begin{prop}[Continuity of $\hat{M}_{N, \infty}$]\label{prop:MEBhatExplicitLowerBound}
    For arbitrary quantum state $\omega$ and a pure target state $\tau$, satisfying $T(\omega, \tau) \leq \epsilon \leq \frac{1}{4}$, we can lower bound $\hat{M}_{N, \infty}(\omega)$ as follows
    \begin{equation}\label{eq:MEBhatExplicitLowerBound}
        \hat{M}_{N, \infty} (\omega) \geq M_{N, \infty}(\tau) - \tilde{\beta}_N \sqrt{\epsilon} \log_2 d - \tilde{b}(\sqrt{\epsilon}),
    \end{equation}
    where $d$ is an upper bound on the dimension of each local subsystem.
\end{prop}

\begin{proof}[Proof of Proposition~\ref{prop:MEBhatExplicitLowerBound}]
    Consider an arbitrary pure state decomposition of $\omega$, $\omega = \sum_i q_i \psi_i$. Define $t_i \coloneqq T(\psi_i, \tau)$. Since both $\psi_i$ and $\tau$ are pure states, using the relationship between the trace distance and the fidelity gives $t_i^2 = 1 - |\braket{\tau}{\psi_i}|^2$. For the variational representation of the trace distance in Eq.~(\ref{eq:TraceDistanceVariational}), by setting $M = \tau$, since $\tau^2 = \tau$ and $\text{Tr}[\tau] = 1$, we have
\begin{equation}
    \begin{split}
        \epsilon \geq T (\omega, \tau)&\geq |\text{Tr}[\tau \omega] - \text{Tr}[\tau^2]|\\
        &= 1 - \text{Tr}[\tau \omega]\\
        &= 1 - \sum_i q_i |\braket{\tau}{\psi_i}|^2\\
        &= \sum_i q_i t_i^2.
    \end{split}
\end{equation}
Consequently, using the Cauchy-Schwarz inequality gives $\bar{t} \coloneqq \sum_i q_i t_i \leq \sqrt{\sum_i q_i t_i^2} \leq \sqrt{\epsilon}$. Then, using Eq.~(\ref{eq:MEBExplicitLowerBound}) gives
\begin{equation}
    \begin{split}
        \sum_i q_i M_{N,\infty}(\psi_i) &\geq \sum_i q_i \{ M_{N, \infty} (\tau)  - \tilde{\beta}_N t_i \log_2 d - \tilde{b}(t_i) \}\\
        &= M_{N, \infty}(\tau) - \tilde{\beta}_N \bar{t} \log_2 d - \sum_i q_i \tilde{b}(t_i).
    \end{split}
\end{equation}
The binary entropy function $h(p)$ is concave on $[0, 1]$ and monotonically increasing on $[0, 1/2]$. In addition, for $p > 0$, $g'(p)  = \log_2 \frac{1+p}{p} > 0$ and $g''(p) = -\frac{1}{(\ln 2) p (1+p) } < 0$. Therefore, $g (p)$ is also concave and monotonically increasing. Hence we have $\sum_i q_i h(t_i) \leq h(\sum_i q_i t_i) = h(\bar{t})$ and $\sum_i q_i g(t_i) \leq g(\bar{t})$. Furthermore, since $\epsilon \leq 1/4$, $\bar{t} \leq \sqrt{\epsilon} \leq 1/2$, resulting in $h(\bar{t}) \leq h(\sqrt{\epsilon})$ and $g(\bar{t}) \leq g(\sqrt{\epsilon})$. Consequently, we have $\sum_i q_i \tilde{b}(t_i) \leq \tilde{b} (\bar{t}) \leq \tilde{b}(\sqrt{\epsilon})$. Therefore,
\begin{equation}
    \begin{split}
        \sum_i q_i M_{N,\infty} (\psi_i) &\geq M_{N, \infty} (\tau) -\tilde{\beta}_N \sqrt{\epsilon} \log_2 d - \tilde{b}(\sqrt{\epsilon}).
    \end{split}
\end{equation}
The above inequality holds for arbitrary pure-state decompositions. Thus, taking the infimum concludes the proof.
\end{proof}

The same argument, together with the continuity bound for $M_{\mathcal{P}, \infty}$, gives
\begin{equation}\label{eq:MEBhatExplicitLowerBoundFixed}
    \hat{M}_{\mathcal{P}, \infty} (\omega) \geq M_{\mathcal{P}, \infty} (\tau) - \beta_m \sqrt{\epsilon} \log_2 d_{\mathcal{P}} - b_m(\sqrt{\epsilon}) 
\end{equation}
for any fixed partition $\mathcal{P}$ with $m = |\mathcal{P}|\geq 2$.

\subsection{5. Upper bound on the distillable GHZ state}
Combining the above results, we first prove that $M_{N, \infty}$ upper bounds the distillable GHZ state for pure states:

\begin{thm}\label{Theorem:Upperbound}
    For an arbitrary finite-dimensional pure $N$-partite state $\psi$,
    \begin{equation}\label{eq:UpperBound}
        D_{\text{GHZ}, N} (\psi) \leq M_{N, \infty} (\psi).
    \end{equation}
\end{thm}

\begin{proof}[Proof of Theorem~\ref{Theorem:Upperbound}]
    Consider an achievable distillation protocol. For a sufficiently large $n$, by the definition of the achievable rate, $\epsilon_n \leq 1/4$. Therefore, combining Eqs.~(\ref{eq:MEBhatLOCCMonotoneLambda}), (\ref{eq:MEBhatWeakAdditivity}), (\ref{eq:MEBhatExplicitLowerBound}) gives
    \begin{equation}
        \begin{split}
            n M_{N, \infty} (\psi) &= \hat{M}_{N, \infty} (\psi^{\otimes n})\\
            &\geq \hat{M}_{N, \infty} (\Lambda_n (\psi^{\otimes n}))\\
            &= \hat{M}_{N, \infty} (\omega_n)\\
            &\geq (1-\tilde{\beta}_N \sqrt{\epsilon_n}) \log_2 d_n - \tilde{b}(\sqrt{\epsilon_n}).
        \end{split}
    \end{equation}
    Since $\epsilon_n \to 0$ as $n \to \infty$, we have $1-\tilde{\beta}_N \sqrt{\epsilon_n}>0$ for sufficiently large $n$. Hence,
    \begin{equation}
        \frac{\log_2 d_n}{n} \leq \frac{M_{N, \infty}(\psi) + \tilde{b}(\sqrt{\epsilon_n})/n}{1 - \tilde{\beta}_N \sqrt{\epsilon_n}},
    \end{equation}
    and consequently
    \begin{equation}
        \limsup_{n \rightarrow \infty} \frac{\log_2 d_n}{n} \leq M_{N, \infty}(\psi).
    \end{equation}
    The above discussion holds for arbitrary achievable distillation protocols. Then,
    \begin{equation}
        D_{\text{GHZ}, N}(\psi) \leq M_{N, \infty}(\psi).
    \end{equation}
\end{proof}




\subsection{6. Proof of Theorem~\ref{thm:MEBupper}}
Finally, we provide the proof of Theorem~\ref{thm:MEBupper}. To this end, we introduce a new quantity $\tilde{M}_{N, \infty}$ which forms an upper bound on the convex-roof extension $\hat{M}_{N, \infty}$:

\begin{dfn}
    For an $N$-partite mixed state $\rho$ and a partition $\mathcal{P} = \{ P_1, \ldots, P_m \}$, the MEB $\tilde{M}_{N, \infty}$ is defined as follows
    \begin{equation}\label{eq:MEBtilde}
        \tilde{M}_{N, \infty}(\rho) \coloneqq \min_{\mathcal{P} : |\mathcal{P}|\geq 2} \tilde{M}_{\mathcal{P}, \infty} (\rho),
    \end{equation}
    where
    \begin{equation}\label{eq:MEBtildeFixed}
    \tilde{M}_{\mathcal{P}, \infty}(\rho) \coloneqq \frac{1}{m}\sum_i S(P_i)\rho - \frac{1}{m(m-1)}\sum_i E_R^\infty (\rho_{\bar{i}}).
    \end{equation}
\end{dfn}

Then, we can prove Theorem~\ref{thm:MEBupper} as follows:

\begin{proof}[Proof of Theorem~\ref{thm:MEBupper}]
    Suppose a quantum state $\omega_n$ is generated from $\rho^{\otimes n}$ and LOCC, such that
    \begin{equation}
        T(\omega_n, \Gamma_{d_n}^{(N)}) \leq \epsilon_n, \quad \epsilon_n \rightarrow 0.
    \end{equation}
    From the LOCC monotonicity (Eq.~\eqref{eq:MEBhatLOCCMonotoneLambda}), we have $\hat{M}_{N, \infty} (\rho^{\otimes n}) \geq \hat{M}_{N, \infty} (\omega_n)$. Also, from Eq.~\eqref{eq:MEBhatExplicitLowerBound}, $\hat{M}_{N, \infty}(\omega_n) \geq (1-\tilde{\beta}_N \sqrt{\epsilon_n}) \log_2 d_n - \tilde{b}(\sqrt{\epsilon_n})$. Therefore,
    \begin{equation}
        \begin{split}
            \frac{1}{n}\hat{M}_{N, \infty}(\rho^{\otimes n}) \geq (1-\tilde{\beta}_N \sqrt{\epsilon_n}) \frac{\log_2 d_n}{n} - \frac{\tilde{b}(\sqrt{\epsilon_n})}{n},
        \end{split}
    \end{equation}
    and consequently
    \begin{equation}
        D_{\text{GHZ}, N} (\rho) \leq \liminf_{n \rightarrow \infty} \frac{1}{n} \hat{M}_{N, \infty} (\rho^{\otimes n}).
    \end{equation}

    Using the concavity of the von Neumann entropy and the convexity of the regularized REE (Eq.\eqref{eq:RegularizedREEConvexity}), for an arbitrary pure-state decomposition $\rho = \sum_j p_j \psi^j$ and a fixed partition $\mathcal{P} = \{ P_1, \ldots, P_m \}$, we have
    \begin{equation}
        \begin{split}
            \tilde{M}_{\mathcal{P}, \infty}(\rho) &= \frac{1}{m} \sum_i S(\rho_{P_i}) - \frac{1}{m(m-1)} \sum_i E_R^\infty (\rho_{\bar{i}})\\
            &\geq \sum_j p_j [\frac{1}{m}  \sum_i S((\psi_j)_{P_i}) - \frac{1}{m(m-1)} \sum_i E_R^\infty((\psi_j)_{\bar{i}})]\\
            &= \sum_j p_j M_{\mathcal{P}, \infty} (\psi_j).
        \end{split}
    \end{equation}
    Taking the infimum on the right-hand side gives $\tilde{M}_{\mathcal{P}, \infty}(\rho) \geq \hat{M}_{\mathcal{P}, \infty} (\rho)$. Hence
    \begin{equation}
        \begin{split}
            \tilde{M}_{N,\infty}(\rho) = \min_{\mathcal{P}:|\mathcal{P}|\geq 2} \tilde{M}_{\mathcal{P}, \infty}(\rho) \geq \min_{\mathcal{P}:|\mathcal{P}|\geq 2} \hat{M}_{\mathcal{P}, \infty}(\rho) \geq \hat{M}_{N,\infty}(\rho),
        \end{split}
    \end{equation}
    where the last inequality follows from Eq.~\eqref{eq:MhatInequality}. On the other hand, due to the additivity of the von Neumann entropy and the weak additivity of the regularized REE (Eq.~\eqref{eq:RegularizedREEWeakAdditivity}),
    \begin{equation}
        \begin{split}
            \tilde{M}_{\mathcal{P}, \infty} (\rho^{\otimes n}) &= \frac{1}{m} \sum_i S(\rho_{P_i}^{\otimes n}) - \frac{1}{m(m-1)} \sum_i E_R^\infty(\rho_{\bar{i}}^{\otimes n})\\
            &= \frac{n}{m} \sum_i S(\rho_{P_i}) - \frac{n}{m(m-1)} \sum_i E_R^\infty (\rho_{\bar{i}})\\
            &= n \tilde{M}_{\mathcal{P}, \infty} (\rho).
        \end{split}
    \end{equation}
    Thus, we have
    \begin{equation}
        \begin{split}
            \tilde{M}_{N,\infty}(\rho^{\otimes n}) = \min_{\mathcal{P} : |\mathcal{P}|\geq 2} \tilde{M}_{\mathcal{P}, \infty}(\rho^{\otimes n}) = n \min_{\mathcal{P}:|\mathcal{P}|\geq 2} \tilde{M}_{\mathcal{P}, \infty}(\rho) = n\tilde{M}_{N, \infty} (\rho).
        \end{split}
    \end{equation}
    Consequently,
    \begin{equation}
        \begin{split}
            D_{\text{GHZ}, N}(\rho) &\leq \liminf_{n \rightarrow \infty} \frac{1}{n} \hat{M}_{N, \infty} (\rho^{\otimes n})\\
            &\leq \liminf_{n \rightarrow \infty} \frac{1}{n} \tilde{M}_{N, \infty} (\rho^{\otimes n})\\
            &= \tilde{M}_{N, \infty} (\rho).
        \end{split}
    \end{equation}
\end{proof}

\section{C. Detailed proof of Theorem~\ref{thm:ProductBounds}}
In this section, we provide the detailed proof of Theorem~\ref{thm:ProductBounds}. To this end, we first consider a Bell-state graph which is a finite graph $G = (V, E)$ where $|V| = N$ and, for each edge $e \in E$, there is a $d_e$-dimensional maximally entangled state $\ket{\Phi_{d_e}} \coloneqq \frac{1}{\sqrt{d_e}} \sum_{j = 0}^{d_e -1} \ket{jj}$. The whole system of this graph can be expressed as $\ket{\Phi_G} \coloneqq \bigotimes_{e \in E} \ket{\Phi_{d_e}}$.

\subsection{1. REE on a Bell-state graph}
We first calculate the multipartite REE and the multipartite regularized REE in Eqs.~\eqref{eq:REE} and \eqref{eq:ReguralizedREE}, respectively, on this Bell-state network:
\begin{lem}\label{lem:REEonBellPairNetwork}
    For the finite graph $G = (V, E)$ where the state $\ket{\Phi_G}$ is distributed, we have
    \begin{equation}\label{eq:REEonBellPairNetwork}
        E_R (\Phi_G) = E_R^\infty (\Phi_G) = \sum_{e \in E} \log_2 d_e.
    \end{equation}
\end{lem}

\begin{proof}[Proof of Lemma~\ref{lem:REEonBellPairNetwork}]
    We first show $E_R \leq \sum_{e \in E} \log_2 d_e$. Consider the following separable state: $\sigma_e = \frac{1}{d_e} \sum_{j = 0}^{d_e -1} \ketbra{jj}{jj}$. Then, $\sigma_G = \bigotimes_{e \in E} \sigma_e$ becomes a fully-separable state. Since $\sigma_e \ket{\Phi_{d_e}} = \frac{1}{d_e}\ket{\Phi_{d_e}}$, the following holds
    \begin{equation}
        \begin{split}
            D(\Phi_{d_e}||\sigma_e) &= \text{Tr}[\Phi_{d_e} \log_2 \Phi_{d_e}] - \text{Tr}[\Phi_{d_e} \log_2 \sigma_e]\\
            &= -\ev{\log_2 \sigma_e}{\Phi_{d_e}}\\
            &= \log_2 d_e,
        \end{split}
    \end{equation}
    where the first equality is the definition of the quantum relative entropy and the second equality holds since $\Phi_{d_e}$ is pure. The additivity of the quantum relative entropy gives
    \begin{equation}
        \begin{split}
            D(\Phi_G || \sigma_G) &= \sum_{e \in E} D(\Phi_{d_e} || \sigma_e) = \sum_{e\in E} \log_2 d_e.
        \end{split}
    \end{equation}
    Consequently, from the definition of the REE,
    \begin{equation}
        E_R(\Phi_G) \leq D(\Phi_G || \sigma_G) = \sum_{e\in E} \log_2 d_e.
    \end{equation}

    We next show $E_R(\Phi_G) \geq \sum_{e\in E} \log_2 d_e$. To this end, we first prove that, for arbitrary fully-separable state $\sigma$, 
    \begin{equation}
        D(\Phi_G || \sigma) \geq \sum_{e\in E} \log_2 d_e.
    \end{equation}
    Let $\ket{a_v}$ denote a normalized state on a local system of party $v \in V$. Define $\ket{a} \coloneqq \bigotimes_{v \in V} \ket{a_v}$. Suppose party $v$ performs a projection measurement $P_v = \ketbra{a_v}{a_v}$. Then, the joint probability of obtaining all these outcomes is
    \begin{equation}
        \begin{split}
            P(a_1, \ldots, a_N) &= \ev{\Phi_G}{\bigotimes_v P_v}{\Phi_G} = |\braket{a}{\Phi_G}|^2.
        \end{split}
    \end{equation}
    Let $D_v$ denote a product of local dimensions of edges connecting party $v$, i.e., $D_v \coloneqq \prod_{e \ni v} d_e$. Then, the reduced state of party $v$ can be expressed as $\rho_v = \frac{1}{D_v}  I_{D_v}$, and therefore $P(a_v) = \text{Tr}[\rho_v P_v] = 1/{D_v}$. We use the following quantum Finner inequality~\cite{Renou2019Limits}: $P(a_1, \ldots, a_N) \leq \prod_{v \in V} P(a_v)^{1/2}$. Then, we have
    \begin{equation}
        \begin{split}
            |\braket{a}{\Phi_G}|^2 &= P(a_1, \ldots, a_N)\\
            &\leq \prod_{v \in V} P(a_v)^{1/2}\\
            &= \prod_{v \in V} D_v^{-1/2}\\
            &= \prod_{v \in V} \prod_{e \ni v} d_e^{-1/2}.
        \end{split}
    \end{equation}
    Here, since each edge $e$ is connected with two vertices, 
    \begin{equation}
        \begin{split}
            \prod_{v \in V} \prod_{e \ni v} d_e^{-1/2} &= \prod_{e \in E } (d_e^{-1/2})^2\\
            &=2^{\log_2 \prod_{e \in E} d_e^{-1}}  \\
            &= 2^{-\sum_{e} \log_2 d_e}
        \end{split}
    \end{equation}
    Consequently, $|\braket{a}{\Phi_G}|^2 \leq 2^{-\sum_{e} \log_2 d_e}$. Here, a fully-separable state $\sigma$ can be expressed as a convex combination of pure product states, i.e., $\sigma = \sum_x p_x \ketbra{a_x}{a_x}$, where $\ket{a_x} = \bigotimes_v \ket{a_{x, v}}$. Thus,
    \begin{equation}
    \begin{split}
        \ev{\sigma}{\Phi_G} &= \sum_x p_x |\braket{a_x }{\Phi_G}|^2 \leq \sum_x p_x 2^{-\sum_{e} \log_2 d_e} = 2^{-\sum_{e} \log_2 d_e}.
    \end{split}
    \end{equation}
    Therefore, the following holds
    \begin{equation}
        \begin{split}
            D(\Phi_G||\sigma) &= - \ev{\log_2 \sigma}{\Phi_G}\\
            &\geq - \log_2 \ev{\sigma}{\Phi_G}\\
            &\geq \sum_{e} \log_2 d_e,
        \end{split}
    \end{equation}
    where the first equality holds since $\Phi_G$ is pure and the inequality follows from the operator concavity of the logarithm. Consequently, we have $E_R(\Phi_G) \geq \sum_{e} \log_2 d_e$, and
    \begin{equation}
        E_R (\Phi_G) = \sum_{e \in E} \log_2 d_e.
    \end{equation}

    For $n$ copies of $\Phi_G$, the local dimension of each edge $e$ becomes $d_e^{n}$. We can use the same argument as the above for this case. Therefore, we have
    \begin{equation}
        E_R(\Phi_G^{\otimes n}) = \sum_{e \in E} \log_2 d_e^n = n \sum_{e} \log_2 d_e.
    \end{equation}
    Finally, the regularized REE of $\Phi_G$ becomes
    \begin{equation}
        E_R^\infty (\Phi_G) = \lim_{n \rightarrow \infty} \frac{1}{n} E_R (\Phi_G^{\otimes n}) = \sum_{e} \log_2 d_e.
    \end{equation}
    This concludes the proof of this lemma.
\end{proof}

\subsection{2. Distillable GHZ in a Bell-state graph}
We now derive the distillable GHZ in the Bell-state graph. We split $N$ parties into $m$ groups, i.e., $\mathcal{P} = \{P_1, \ldots, P_m  \}$. Let $E (\mathcal{P})$ denote a set of edges connecting different groups. Then, we have the following theorem:

\begin{thm}\label{Theorem:BellpairNetwork}
    For an $N$-partite Bell-state graph $G$, the distillable GHZ rate is given by
    \begin{equation}\label{eq:BellpairNetwork}
        D_{\text{GHZ}, N} (\Phi_G) = \min_{\mathcal{P} : |\mathcal{P}| \geq 2} \frac{\sum_{e \in E(\mathcal{P})} \log_2 d_e}{|\mathcal{P}|-1},
    \end{equation}
    where the minimum is taken over all partitions $\mathcal{P}$ of the $N$ parties into at least two blocks.
\end{thm}

\begin{proof}[Proof of Theorem~\ref{Theorem:BellpairNetwork}]
    Consider a reduced state of group $P_i$. There exist maximally mixed states for edges connected with the other groups. Hence
    \begin{equation}
        S(P_i) = \sum_{\substack{e \in E(\mathcal{P})\\ e \ni P_i}} \log_2 d_e,
    \end{equation}
    where edges contained in the group $P_i$ do not affect the entropy. When we consider the sum of the von Neumann entropy of each group, since each edge is counted twice, we have $\sum_{i = 1}^m S(P_i) = 2 \sum_{e \in E(\mathcal{P})} \log_2 d_e$. On the other hand, when we trace out the group $P_i$ from $\Phi_G$, there exist local maximally mixed states on edges connected to $P_i$ and Bell states on the other edges are kept. Then, using Eq.~\eqref{eq:REEonBellPairNetwork} gives
    \begin{equation}
        E_{\bar{i}} (\Phi_G) = \sum_{\substack{e \in E (\mathcal{P})\\
        P_i \notin e}} \log_2 d_e.
    \end{equation}
    Considering the sum of the REE, since each edge is counted $m-2$ times, we have $\sum_{i = 1}^m E_{\bar{i}} (\Phi_G) = \sum_{e \in E(\mathcal{P})} (m-2) \log_2 d_e$. Therefore,
    \begin{equation}
        \begin{split}
            M_{\mathcal{P}, \infty} (\Phi_G) &= \frac{1}{m} \sum_{i = 1}^m S(P_i) - \frac{1}{m(m-1)} \sum_{i = 1}^m E_{\bar{i}} (\Phi_G)= \frac{1}{m-1} \sum_{e \in E(\mathcal{P})} \log_2 d_e.
        \end{split}
    \end{equation}
    Consequently, from Eq.~\eqref{eq:UpperBound} 
    \begin{equation}
        D_{\text{GHZ}, N} (\Phi_G) \leq \min_{\mathcal{P} : |\mathcal{P}| \geq 2} \frac{\sum_{e \in E(\mathcal{P})} \log_2 d_e}{|\mathcal{P}| -1}.
    \end{equation}

    In the following, we show $D_{\text{GHZ}, N} (\Phi_G) \geq \min_{\mathcal{P} : |\mathcal{P}| \geq 2} (\sum_{e \in E(\mathcal{P})} \log_2 d_e)/(|\mathcal{P} |-1)$. To this end, we utilize the fractional spanning-tree packing. A spanning tree with $N$ vertices has $N-1$ edges. In addition, we can generate an $N$-partite GHZ state from $N-1$ copies of Bell states deterministically with LOCC, that is, we can generate a GHZ state from a spanning tree. Here, the optimal rate of fractional spanning-tree packing in the finite graph $G$ can be expressed as~\cite{Tutte1961On, NashWilliams1961Edge}
    \begin{equation}
        \min_{\mathcal{P} : |\mathcal{P}| \geq 2} \frac{\sum_{e \in E(\mathcal{P})} \log_2 d_e}{|\mathcal{P} | -1}.
    \end{equation}
    Therefore, we have the following lower bound on the distillable GHZ rate
    \begin{equation}
        D_{\text{GHZ}, N} (\Phi_G) \geq \min_{\mathcal{P} : |\mathcal{P}| \geq 2} \frac{\sum_{e \in E(\mathcal{P})} \log_2 d_e}{|\mathcal{P}| -1}.
    \end{equation}
    This concludes the proof.
\end{proof}

\subsection{3. Proof of Theorem~\ref{thm:ProductBounds}}
Finally, we describe the proof of Theorem~\ref{thm:ProductBounds}. We first recall Theorem~\ref{thm:ProductBounds} in the main text:
\begingroup
\renewcommand{\thethm}{\ref{thm:ProductBounds}}

\begin{thm}
The distillable GHZ rate of an $N$-partite finite graph with a state $\bigotimes_e \rho_e$ is bounded as
    \begin{equation}\label{appeq:MixedNetwork}
    \begin{split}
        \min_{\mathcal{P} : |\mathcal{P} |\geq 2} \frac{\sum_{e \in E(\mathcal{P})} E_D(\rho_e)}{|\mathcal{P}|-1} &\leq D_{\text{GHZ}, N}(\bigotimes_e \rho_e)\leq \min_{\mathcal{P}:|\mathcal{P}|\geq 2} \frac{\sum_{e \in E(\mathcal{P})} E_C(\rho_e)}{|\mathcal{P}|-1},
    \end{split}
    \end{equation}
    where $\mathcal{P}$ denotes the partitioning of the parties and $E(\mathcal{P})$ denotes a set of edges connecting different blocks. In particular, when $E_C (\rho_e) = E_D (\rho_e)$, we have
    \begin{equation}
        D_{\text{GHZ}, N}(\bigotimes_e \rho_e)
        = \min_{\mathcal{P}:|\mathcal{P}|\geq 2} \frac{\sum_{e \in E(\mathcal{P})} E_D(\rho_e)}{|\mathcal{P}|-1}.
    \end{equation}
\end{thm}

\endgroup
\addtocounter{thm}{-1}
Here, $E_D(\rho_e)$ and $E_C (\rho_e)$ denote distillable entanglement and entanglement cost for $\rho_e$. We first consider the lower bound. From many copies of $\rho_e$, we distil several copies of maximally entangled states at rate $E_D(\rho_e)$. From these maximally entangled states, we can generate GHZ states via LOCC~\cite{Meignant2019Distributing}. The maximal achievable GHZ-state-generation rate in this setting is given by the Tutte--Nash-Williams theorem~\cite{Tutte1961On, NashWilliams1961Edge}. Consequently, we obtain the lower bound on the distillable GHZ rate. The upper bound can be obtained as follows. Consider a Bell-state graph where, for each edge $e \in E$, there is a $d_e$-dimensional maximally entangled state $\ket{\Phi_{d_e}} \coloneqq \frac{1}{\sqrt{d_e}} \sum_{j = 0}^{d_e-1} \ket{jj}$, and a state of the whole system is $\ket{\Phi_G} = \bigotimes_{e \in E} \ket{\Phi_{d_e}}$. By the definition of the entanglement cost, $n$ copies of $\rho_e$ can be prepared from $n E_C(\rho_e)$ Bell states with vanishing error via LOCC. Then, using Theorem~\ref{Theorem:BellpairNetwork} gives the upper bound on the distillable GHZ rate, concluding the proof.

\section{D. Distillable GHZ for a product of bipartite states}
Suppose that three parties, $A$, $B$, and $C$, share the state $\ket{\psi}_{ABC} = \ket{\Phi}_{AB}^{\otimes a} \ket{\Phi}_{AC}^{\otimes b} \ket{\Phi}_{BC}^{\otimes c}$, where $\ket{\Phi}$ is the Bell state and $a$, $b$, and $c$ are some positive integers. In this scenario, we can consider the following four partitionings
\begin{equation}
    \mathcal{P}_1 = \{A, B, C  \}, \quad \mathcal{P}_2 = \{A, BC  \}, \quad \mathcal{P}_3 = \{B, AC  \}, \quad \mathcal{P}_4 = \{C, AB  \}.
\end{equation}
The MEB with respect to each partition is
\begin{equation}
    M_{\mathcal{P}_1, \infty}(\psi) = \frac{a+b+c}{2}, \quad M_{\mathcal{P}_2, \infty}(\psi) = a+b, \quad  M_{\mathcal{P}_3, \infty}(\psi) = a + c, \quad M_{\mathcal{P}_4, \infty}(\psi) = b + c.
 \end{equation}
Therefore, by Theorem~\ref{thm:ProductBounds}, the GHZ-distillation rate is given by
\begin{equation}
    D_{\text{GHZ}, 3} (\psi) = \min \{\frac{a+b+c}{2}, a+b, a+c, b+c  \}.
\end{equation}
When $a = b = c = 1$, we have $D_{\text{GHZ}, 3}(\psi) = 3/2$.

\section{E. Proof of Theorem 5}
Consider a specific partition $\bar{\mathcal{P}}$ of the $N$ parties into at least two blocks, each containing at least one of the $S$ target parties. Then, for each block, all parties in the same block can perform joint operations on their systems. In other words, each block is regarded as one big party. Let $Q^\leftrightarrow_{\text{GHZ}} (\bar{\mathcal{P}})$ denote the GHZ distribution capacity of the quantum network after the partition. Then, clearly we have $Q_{\text{GHZ}, S}^\leftrightarrow (\{\mathcal{N}_e \}) \leq Q^\leftrightarrow_{\text{GHZ}} (\bar{\mathcal{P}})$ since any possible operations in the original network can be realized in the scenario after the partition. In addition, by teleportation simulation and the upper bound in Theorem~\ref{thm:ProductBounds}, we observe that
\begin{equation}
    Q^\leftrightarrow_{\text{GHZ}} (\bar{\mathcal{P}}) \leq \frac{\sum_{e \in E(\bar{\mathcal{P}})} E_C(\omega_e)}{|\bar{\mathcal{P}}|-1}.
\end{equation}
Therefore, we have
\begin{equation}
    Q_{\text{GHZ}, S}^\leftrightarrow (\{\mathcal{N}_e \}) \leq \min_{\bar{\mathcal{P}} : |\bar{\mathcal{P}}| \geq 2} \frac{\sum_{e \in E(\bar{\mathcal{P}})} E_C(\omega_e)}{|\bar{\mathcal{P}}|-1}.
\end{equation}
Since this holds for any such partition, we obtain Theorem~\ref{thm:Partial}.

\section{F. Common key distribution}
In this section, we consider common key distribution in quantum networks, i.e., parties aim to establish a common secret key. Parties try to generate a multipartite private state $\gamma$ to generate a common key. Let party $A_i$ hold key register $K_{A_i}$ and shield $S_{A_i}$. Let $K = K_{A_1} \cdots K_{A_N}$ and $S = S_{A_1} \cdots S_{A_N}$. Then, a multipartite private state with $d$-dimensional key registers can be expressed as~\cite{Augusiak2009Multipartite}
\begin{equation}\label{eq:MultipartitePrivateState}
    \gamma_{KS} = U (\Gamma_d^{(N)} \otimes \tau_S) U^\dagger,
\end{equation}
where $U$ is a twisting unitary operator. Let $\boldsymbol{j} = (j_1, \ldots, j_N)$ label the computational basis of the key register $K$. Then, the twisting unitary operator has the form
\begin{equation}\label{eq:TwistingUnitary}
    U= \sum_{\boldsymbol{j} \in \{ 0, \ldots, d-1 \}^N} \ketbra{\boldsymbol{j}}{\boldsymbol{j}}_K \otimes U_{\boldsymbol{j}},
\end{equation}
where each $U_{\boldsymbol{j}}$ is a unitary operator on $S$.

We first prove the following lemma:
\begin{lem}\label{Lemma:MEBLowerBoundMI}
    For a fixed partition $\mathcal{P} = \{  P_1, \ldots, P_m\}$, we select one representative party from each block $P_i$ and denote the key register of this party by $K_{P_i}$. The corresponding measurement outcome in the computational basis is denoted by $X_i$. Then, for an arbitrary pure state $\psi$, we have
    \begin{equation}\label{eq:MEBLowerBoundMI}
        M_{\mathcal{P}, \infty}(\psi) \geq \frac{1}{m-1} \sum_{i \neq k} I(X_k : X_i),
    \end{equation}
    where $I(X_k: X_i)$ denotes the mutual information between $X_k$ and $X_i$ for $i \neq k$.
\end{lem}

\begin{proof}[Proof of Lemma~\ref{Lemma:MEBLowerBoundMI}]
    In the proof of Lemma~\ref{lem:PureMEBLOCCMonotone}, we prove the following inequality
    \begin{equation}
        M_{\mathcal{P}, \infty} (\psi) - \sum_x p_x M_{\mathcal{P}, \infty} (\psi^x) \geq \frac{1}{m-1} \sum_{i \neq k} [S(P_i)_\psi - \sum_x p_x S(P_i)_{\psi^x}],
    \end{equation}
    where $\psi^x$ is a pure quantum state after one party performs a local quantum instrument and $p_x$ is the probability of obtaining outcome $x$. After the measurement of block $P_k$, a reduced state between $P_k$ and $P_i$ can be expressed as the classical-quantum state $\rho_{X_kP_i} = \sum_{x_k} p_{x_k} \ketbra{x_k}{x_k} \otimes \psi_{P_i}^{x_k}$ where $\psi_{P_i}^{x_k}$ is a conditional quantum state given that $P_k$'s measurement outcome is $x_k$. Then, by the definition of the mutual information,
    \begin{equation}
        \begin{split}
            I(X_k : P_i) &= H(X_k) + S(P_i) - S(X_k P_i)\\
            &= H(X_k) + S(P_i) - (H(X_k) + \sum_{x_k} p_{x_k} S(\psi_{P_i}^{x_k}))\\
            &= S(P_i) - \sum_{x_k} p_{x_k} S(\psi_{P_i}^{x_k}).
        \end{split}
    \end{equation}
    Using the data-processing inequality gives $I(X_k : P_i) \geq I(X_k : X_i)$. In addition, due to the nonnegativity of the MEB (Eq.~\eqref{eq:MEBNonnegativityFixed}), we have
    \begin{equation}
        \begin{split}
            M_{\mathcal{P}, \infty} (\psi) &\geq M_{\mathcal{P}, \infty}(\psi) - \sum_x p_x M_{\mathcal{P}, \infty} (\psi^x)\\
            &\geq \frac{1}{m-1} \sum_{i \neq k} I(X_k : P_i)\\
            &\geq \frac{1}{m-1} \sum_{i \neq k} I(X_k : X_i).
        \end{split}
    \end{equation}
\end{proof}

With this lemma, we prove the following proposition:
\begin{prop}\label{Prop:MEBhatLowerBoundPrivate}
    Consider a fixed partition $\mathcal{P} = \{  P_1, \ldots, P_m\}$. For a multipartite private state $\gamma$ with $d$-dimensional key registers and a quantum state $\omega$ satisfying $T(\omega, \gamma) \leq \epsilon \leq 1/4$,
    \begin{equation}\label{eq:MEBhatLowerBoundPrivate}
        \hat{M}_{\mathcal{P},\infty} (\omega) \geq (1-\sqrt{\epsilon} - \epsilon) \log_2 d - h(\sqrt{\epsilon}) - h(\epsilon).
    \end{equation}
\end{prop}

\begin{proof}[Proof of Proposition~\ref{Prop:MEBhatLowerBoundPrivate}]
    Consider an arbitrary pure-state decomposition of $\omega$, $\omega = \sum_z q_z \psi^z$. By the definition, $\hat{M}_{\mathcal{P}, \infty} (\omega) = \inf_{\omega = \sum_z q_z \psi^z} \sum_z q_z M_{\mathcal{P}, \infty}(\psi^z)$. Using Lemma~\ref{Lemma:MEBLowerBoundMI} gives
    \begin{equation}
        \begin{split}
            \sum_z q_z M_{\mathcal{P}, \infty}(\psi^z) &\geq \frac{1}{m-1} \sum_{i \neq k} \sum_z q_z I(X_k : X_i)_{\psi^z} = \frac{1}{m-1} \sum_{i \neq k} I(X_k : X_i|Z),
        \end{split}
    \end{equation}
    where $Z$ is a classical register representing the labels of the pure-state decomposition. From the definition of the conditional mutual information, we have $I(X_k: X_i|Z) = H(X_k|Z) - H(X_k | X_iZ)$. Therefore,
    \begin{equation}\label{eq:AverageLowerBound}
    \begin{split}
        \sum_z q_z M_{\mathcal{P}, \infty}(\psi^z) &\geq \frac{1}{m-1} \sum_{i \neq k} (H(X_k|Z) - H(X_k|X_iZ))\\
        &= H(X_k|Z) - \frac{1}{m-1} \sum_{i \neq k} H(X_k|X_i Z).
    \end{split}
    \end{equation}

    We first derive a lower bound on the conditional entropy $H(X_k|Z)$. To this end, we follow the privacy test approach of Ref.~\cite{Wilde2017Converse}. For the multipartite private state in Eq.~\eqref{eq:MultipartitePrivateState}, we consider the following privacy test
    \begin{equation}
        \Pi_\gamma \coloneqq U(\Gamma_d^{(N)} \otimes I_S) U^\dagger.
    \end{equation}
    The multipartite private state passes this privacy test with unit probability, i.e., $\text{Tr}[\Pi_\gamma \gamma] = 1$.
    In addition, due to the variational representation of the trace distance,
    \begin{equation}
        \begin{split}
            \epsilon &\geq T(\omega, \gamma) \geq |\text{Tr}[\Pi_\gamma \omega] - \text{Tr}[\Pi_\gamma \gamma]|= 1 - \text{Tr}[\Pi_\gamma \omega].
        \end{split}
    \end{equation}
    That is, $\text{Tr}[\Pi_\gamma \omega] \geq 1 - \epsilon$. Here, for the pure-state decomposition $\omega = \sum_z q_z \psi^z$, define $f_z \coloneqq \text{Tr}[\Pi_\gamma \psi^z]$ which corresponds to the probability that $\psi^z$ passes the privacy test. Then,
    \begin{equation}\label{eq:AvarageDistancePrivate}
        \begin{split}
            \sum_z q_z (1-f_z) &= 1 - \sum_z q_z \text{Tr}[\Pi_\gamma \psi^z]= 1 - \text{Tr}[\Pi_\gamma \omega]\leq \epsilon.
        \end{split}
    \end{equation}
    Suppose $f_z > 0$ and consider the quantum state $\ket{\varphi^z} \coloneqq \frac{1}{\sqrt{f_z}} \Pi_\gamma \ket{\psi^z}$. From the definition of the privacy test, $(\Gamma_d^{(N)} \otimes I_S)U^\dagger \ket{\varphi^z} = U^\dagger \ket{\varphi^z}$. Therefore, $U^\dagger \ket{\varphi^z} = \ket{\Gamma_d^{(N)}} \otimes \ket{s^z}$ for a normalized shield state $\ket{s^z}$. Hence, we can rewrite $\ket{\varphi^z}$ as $\ket{\varphi^z} = U(\ket{\Gamma_d^{(N)}} \otimes \ket{s^z})$. Since the twisting unitary can be expressed as in Eq.~\eqref{eq:TwistingUnitary},
    \begin{equation}
        \ket{\varphi^z} = \frac{1}{\sqrt{d}} \sum_{j = 0}^{d-1} \ket{j \cdots j}_K \otimes U_{(j, \ldots, j)} \ket{s^z}_S.
    \end{equation}
    Therefore, when we measure the key registers of $\ket{\varphi^z}$, measurement outcomes will be uniformly distributed, i.e., $P(X_1 = \cdots = X_m = j) = \frac{1}{d}$ for $j = 0, \ldots, d-1$. Moreover, since both $\ket{\psi^z}$ and $\ket{\varphi^z}$ are pure states, a trace distance between these two states is $T(\psi^z, \varphi^z) = \sqrt{1 - |\braket{\psi^z}{\varphi^z}|^2}= \sqrt{1 - f_z}$. Let $P_z$ represent a probability distribution of measurement outcomes when block $P_k$ measures its key register of $\psi^z$ and $U_d$ represents the uniform distribution. Then, define a trace distance between these two probability distributions as $t_z \coloneqq \frac{1}{2} \sum_j |P_z (j) - \frac{1}{d}|$. Due to the monotonicity of the trace distance, we have $t_z \leq \sqrt{1-f_z}$. Hence
    \begin{equation}
        \begin{split}
            \bar{t} \coloneqq \sum_z q_z t_z &\leq \sum_z q_z \sqrt{1-f_z}\\
            &\leq \sqrt{\sum_z q_z (1-f_z)}\\
            &\leq \sqrt{\epsilon},
        \end{split}
    \end{equation}
    where the second inequality follows from the concavity of a square-root function and the last inequality follows from Eq.~\eqref{eq:AvarageDistancePrivate}. Using the continuity of the entropy (Eq.~\eqref{eq:VonNeumannContinuity}) gives $|H(P_z) - H(U_d)| \leq t_z \log_2 d + h(t_z)$, and consequently $H(P_z) \geq (1-t_z) \log_2 d - h(t_z)$. That is,
    \begin{equation}
        H(X_k|Z=z) \geq (1-t_z) \log_2 d - h(t_z).
    \end{equation}
    Therefore,
    \begin{equation}
        \begin{split}
            H(X_k|Z) &= \sum_z q_z H(X_k|Z = z)\\
            &\geq (1-\bar{t}) \log_2 d - \sum_z q_z h(t_z)\\
            &\geq (1-\bar{t}) \log_2 d - h(\bar{t}),
        \end{split}
    \end{equation}
    where we used the concavity of the binary entropy function for the last inequality. Furthermore, since $\bar{t} \leq \sqrt{\epsilon} \leq 1/2$ and $h(t)$ is monotonically increasing on $[0, 1/2]$,
    \begin{equation}\label{eq:HXZLowerBound}
        H(X_k|Z) \geq (1-\sqrt{\epsilon})\log_2 d - h(\sqrt{\epsilon}).
    \end{equation}
    When $f_z = 0$, clearly we have $t_z \leq 1 = \sqrt{1-f_z}$ and the same lower bound holds. 

    Next, we derive an upper bound on the conditional entropy $H(X_k|X_iZ)$. For the multipartite private state $\gamma$, measurement outcomes of the key registers of all parties perfectly coincide, i.e., $P_\gamma (X_k \neq X_i) = 0$ for all $i \neq k$. On the other hand, for the quantum state $\omega$ satisfying $T(\omega, \gamma) \leq \epsilon$, we have $p_i \coloneqq P_\omega (X_k \neq X_i) \leq \epsilon$. Using Fano's inequality gives $H(X_k | X_i) \leq h(p_i) + p_i \log_2 d$. Moreover, since $\epsilon \leq 1/4$ and $H(X_k | X_i Z) \leq H(X_k | X_i)$, we have
    \begin{equation}\label{eq:HXXZUpperBound}
        H(X_k | X_i Z) \leq h (\epsilon) + \epsilon \log_2 d.
    \end{equation}
    Combining Eqs.~\eqref{eq:AverageLowerBound}, \eqref{eq:HXZLowerBound}, and \eqref{eq:HXXZUpperBound} gives
    \begin{equation}
        \begin{split}
            \sum_z q_z M_{\mathcal{P}, \infty}(\psi^z) &\geq H(X_k | Z) - \frac{1}{m-1} \sum_{i \neq k} H (X_k | X_i Z)\\
            &\geq (1-\sqrt{\epsilon}) \log_2 d - h(\sqrt{\epsilon}) - \frac{1}{m-1}\sum_{i \neq k} (h(\epsilon) + \epsilon \log_2 d)\\
            &= (1 - \sqrt{\epsilon} - \epsilon) \log_2 d - h(\sqrt{\epsilon}) - h(\epsilon).
        \end{split}
    \end{equation}
    Taking the infimum of the left-hand side gives
    \begin{equation}
        \hat{M}_{\mathcal{P}, \infty}(\omega) \geq (1-\sqrt{\epsilon}-\epsilon) \log_2 d - h(\sqrt{\epsilon}) - h(\epsilon).
    \end{equation}
\end{proof}

Suppose we generate a quantum state $\omega_n$ from $n$ copies of a quantum state $\rho$ using LOCC such that
\begin{equation}\label{eq:DefinitionDistillableKey}
    T(\omega_n, \gamma_{d_n}) \leq \epsilon_n, \quad \epsilon_n \rightarrow 0,
\end{equation}
where $\gamma_{d_n}$ is a multipartite private state whose key dimension is $d_n$. Then, the distillable common-key rate is defined as
\begin{equation}
    K_{\text{c}, N} (\rho) \coloneqq \sup \{ \liminf_{n\rightarrow \infty}  \frac{\log_2 d_n}{n}  \},
\end{equation}
where the supremum is taken over all sequences of LOCC protocols and multipartite private target states satisfying Eq.~\eqref{eq:DefinitionDistillableKey}.

The MEB upper bounds the distillable common-key rate, that is, we have the following theorem:
\begin{thm}\label{Theorem:DistillableCommonKeyUpperBound}
    For an $N$-partite finite-dimensional quantum state $\rho$,
    \begin{equation}
        K_{\text{c}, N} (\rho) \leq \tilde{M}_{N, \infty}(\rho).
    \end{equation}
    In particular, for a pure state $\psi$,
    \begin{equation}
        K_{\text{c}, N}(\psi) \leq M_{N, \infty} (\psi).
    \end{equation}
\end{thm}

\begin{proof}[Proof of Theorem~\ref{Theorem:DistillableCommonKeyUpperBound}]
    Consider a fixed partition $\mathcal{P} = \{ P_1, \ldots, P_m \}$. By the fixed-partition version of Proposition~\ref{Prop:MEBhatLOCCMonotone}, $\hat{M}_{\mathcal{P}, \infty} (\rho^{\otimes n}) \geq \hat{M}_{\mathcal{P}, \infty} (\omega_n)$. Together with Proposition~\ref{Prop:MEBhatLowerBoundPrivate}, we have
    \begin{equation}
        \frac{1}{n}\hat{M}_{\mathcal{P}, \infty} (\rho^{\otimes n}) \geq (1-\sqrt{\epsilon_n} - \epsilon_n) \frac{\log_2 d_n}{n} - \frac{h(\sqrt{\epsilon_n}) + h(\epsilon_n)}{n}.
    \end{equation}
    Since $\epsilon_n \to 0$ as $n \rightarrow \infty$, the error terms vanish in the asymptotic limit. Hence
    \begin{equation}
        K_{\text{c}, N} (\rho) \leq \liminf_{n \rightarrow \infty} \frac{1}{n} \hat{M}_{\mathcal{P}, \infty}(\rho^{\otimes n}).
    \end{equation}
    Using the same argument as in the proof of Theorem~\ref{thm:MEBupper} gives
    \begin{equation}
        K_{\text{c}, N} (\rho) \leq \tilde{M}_{\mathcal{P}, \infty}(\rho).
    \end{equation}
    Since this holds for arbitrary partition, we have
    \begin{equation}
        K_{\text{c}, N} (\rho) \leq \min_{\mathcal{P}:|\mathcal{P}|\geq 2} \tilde{M}_{\mathcal{P}, \infty}(\rho) = \tilde{M}_{N, \infty} (\rho).
    \end{equation}
    For a pure state $\psi$, we have $\tilde{M}_{N,\infty} (\psi) = M_{N, \infty}(\psi)$ by definition. Hence
    \begin{equation}
        K_{\text{c}, N}(\psi) \leq M_{N, \infty}(\psi).
    \end{equation}
\end{proof}

We consider common-key distillation in an $N$-partite finite graph $G = (V, E)$ where each edge $e \in E$ has a $d_e$-dimensional maximally entangled state $\ket{\Phi_{d_e}} = \frac{1}{\sqrt{d_e}} \sum_{j = 0}^{d_e - 1} \ket{jj}$. A quantum state of the whole system is $\ket{\Phi_G} = \bigotimes_{e \in E} \ket{\Phi_{d_e}}$. Consider partitioning $\mathcal{P}$ of $N$ parties. Let $E(\mathcal{P})$ denote a set of edges connecting different groups. The distillable common-key rate after the grouping is at least as high as that before the grouping since parties in the same group can perform joint operations on their systems. 

Now, we have the following theorem:

\begin{thm}\label{Theorem:CommonKeyDistillationBellPairNetwork}
    For an $N$-partite finite graph $G$ with a whole state $\Phi_G$,
    \begin{equation}
        K_{\text{c}, N}(\Phi_G) = \min_{\mathcal{P}:|\mathcal{P}| \geq 2} \frac{\sum_{e \in E(\mathcal{P})} \log_2 d_e}{|\mathcal{P}|-1}.
    \end{equation}
\end{thm}

\begin{proof}[Proof of Theorem~\ref{Theorem:CommonKeyDistillationBellPairNetwork}]
    By using Theorem~\ref{Theorem:DistillableCommonKeyUpperBound} and Lemma~\ref{lem:REEonBellPairNetwork}, we have
    \begin{equation}
        K_{\text{c}, N} (\Phi_G) \leq \min_{\mathcal{P}:|\mathcal{P}| \geq 2} \frac{\sum_{e \in E(\mathcal{P})} \log_2 d_e}{|\mathcal{P}|-1}.
    \end{equation}
    On the other hand, by performing the fractional spanning-tree packing,
    \begin{equation}
        \min_{\mathcal{P}:|\mathcal{P}| \geq 2} \frac{\sum_{e \in E(\mathcal{P})} \log_2 d_e}{|\mathcal{P}|-1} \leq D_{\text{GHZ}, N} (\Phi_G).
    \end{equation}
    In addition, since the parties can generate a common key from one copy of GHZ states, $D_{\text{GHZ}, N} (\Phi_G) \leq K_{\text{c}, N} (\Phi_G)$. Therefore,
    \begin{equation}
        K_{\text{c}, N} (\Phi_G) = D_{\text{GHZ}, N} (\Phi_G) = \min_{\mathcal{P}:|\mathcal{P}| \geq 2} \frac{\sum_{e \in E(\mathcal{P})} \log_2 d_e}{|\mathcal{P}|-1}.
    \end{equation}
\end{proof}

Then, using the same argument as in the proof of Theorem~\ref{thm:ProductBounds}, we derive the following theorem:

\begin{thm}
    The distillable common-key rate of an $N$-partite finite graph with a state $\bigotimes_e \rho_e$ is bounded as
    \begin{equation}
        \min_{\mathcal{P}: |\mathcal{P}|\geq 2} \frac{\sum_{e \in E(\mathcal{P})} E_D(\rho_e)}{|\mathcal{P}|-1} \leq K_{\text{c}, N} (\bigotimes_e \rho_e) \leq \min_{\mathcal{P}: |\mathcal{P}|\geq 2} \frac{\sum_{e \in E(\mathcal{P})} E_C(\rho_e)}{|\mathcal{P}|-1},
    \end{equation}
    where $\mathcal{P}$ denotes the partitioning of the parties. In particular, when $E_C(\rho_e) = E_D(\rho_e)$, we have
    \begin{equation}
        K_{\text{c}, N}(\bigotimes _e \rho_e) = \min_{\mathcal{P}: |\mathcal{P}|\geq 2} \frac{\sum_{e \in E(\mathcal{P})} E_D(\rho_e)}{|\mathcal{P}|-1}.
    \end{equation}
\end{thm}

As in Theorem~\ref{thm:Teleportation}, we can prove the following theorem on the common private capacity in a teleportation-simulable-channel network $\{\mathcal{N}_e  \}$ when we can obtain one copy of a resource state $\omega_e$ by using a teleportation-simulable channel $\mathcal{N}_e$ once and LOCC without preshared entanglement:
\begin{thm}\label{Theorem:CommmonKeyTeleportationsimulable}
    Consider a teleportation-simulable-channel network $\{ \mathcal{N}_e \}$ with resource states $\omega_e$. Suppose that each resource state $\omega_e$ can be obtained from one use of $\mathcal{N}_e$ and LOCC without preshared entanglement. Then, the common private capacity is bounded as follows:
    \begin{equation}
        \min_{\mathcal{P}:|\mathcal{P}|\geq 2} \frac{\sum_{e \in E(\mathcal{P})} E_D (\omega_e)}{|\mathcal{P}|-1} \leq P_{\text{c}, N}^\leftrightarrow (\{ \mathcal{N}_e \})  \leq \min_{\mathcal{P} : |\mathcal{P}| \geq 2} \frac{\sum_{e \in E(\mathcal{P})} E_C(\omega_e)}{|\mathcal{P}|-1}.
    \end{equation}
    In particular, when $E_C(\omega_e) = E_D(\omega_e)$, we have
    \begin{equation}
        P_{\text{c}, N}^\leftrightarrow (\{ \mathcal{N}_e \})  = \min_{\mathcal{P} : |\mathcal{P}| \geq 2} \frac{\sum_{e \in E(\mathcal{P})} E_D(\omega_e)}{|\mathcal{P}|-1}.
    \end{equation}
\end{thm}

For the erasure channel $\mathcal{N}_e^{p_e}$ with the erasure probability $p_e$, $E_D(\omega_e) = E_C(\omega_e) = 1-p_e$ where $\omega_e$ is the corresponding resource state. Therefore, we have the following theorem:

\begin{thm}
    The common private capacity of the quantum network $\{ \mathcal{N}_e^{p_e} \}$ consisting of point-to-point qubit erasure channels is given by
    \begin{equation}
        P_{\text{c}, N}^\leftrightarrow (\{ \mathcal{N}_e^{p_e} \}) = \min_{\mathcal{P} : |\mathcal{P}|\geq 2} \frac{\sum_{e \in E(\mathcal{P})} (1-p_e)}{|\mathcal{P}|-1},
    \end{equation}
    where $\mathcal{N}_e^{p_e}$ represents an erasure channel on edge $e$ with erasure probability $p_e$.
\end{thm}

Furthermore, using the same argument as in the proof of Theorem~\ref{thm:Partial}, we obtain the following theorem:

\begin{thm}
    Suppose, in the $N$-partite teleportation-simulable network $\{ \mathcal{N}_e \}$, common keys are distributed over only $S < N$ parties. Let $\bar{\mathcal{P}}$ be a partitioning of all $N$ parties into at least two blocks, each containing at least one of the $S$ target parties. Then, the common private capacity over $S$ parties is bounded by
    \begin{equation}
        P_{\text{c}, S}^\leftrightarrow (\{  \mathcal{N}_e\}) \leq \min_{\bar{\mathcal{P}} : |\bar{\mathcal{P}}| \geq 2} \frac{\sum_{e \in E(\bar{\mathcal{P}})} E_C(\omega_e)}{|\bar{\mathcal{P}}|-1}.
    \end{equation}
\end{thm}

Note that a quantum state of the whole system after the grouping can still be considered as a multipartite private state. For each group, parties choose one key register and subtract its value from the other registers modulo $d$. This operation can be considered as a local unitary operation in a group. Then, a quantum state on key registers $\ket{j \cdots j}$ becomes $\ket{j} \otimes \ket{0 \cdots 0}$. The state $\ket{0 \cdots 0}$ can be included in a shield system. Then, the whole state after the grouping can be considered as a multipartite private state where each group holds one key register.

Finally, we note that the above theorems hold not only for a full quantum network, where network nodes are supposed to process quantum signals coherently, but also for the trusted-node-based QKD network, where QKD links are terminated at each node, which is assumed to be trusted, and keys are classically relayed to reach arbitrary nodes. The statement is justified as follows: the lower bounds are derived by considering a classical version of the LOCC protocol in Theorem~\ref{thm:Teleportation}. On the other hand, the key relay protocols in the trusted-node-based QKD network are usually classical but one can always interpret them as fully quantum key relays. Then, since the rate of the full quantum version of the key relay (and QKD links) is upper bounded by the above theorems, the rate of the classical key relay with QKD links is also limited by the same bound.

\begin{figure}[tbp]
\centering
\includegraphics[keepaspectratio, scale=0.6]{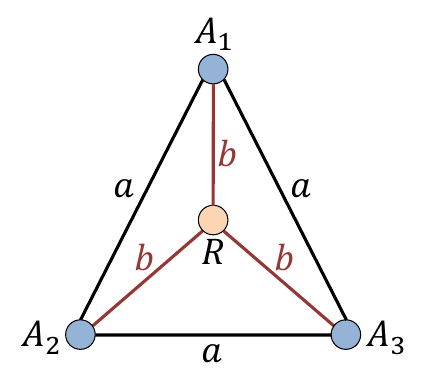}
\caption{\label{fig:RepeaterNetwork}
Schematic of the triangle quantum network with the network quantum repeater node $R$.}
\end{figure}

\section{G. Network quantum repeater}
\subsection{1. Proof of Theorem~\ref{thm:repeater}}
We consider an $N$-partite complete graph $K_N$ with a network quantum repeater node $R$, as illustrated for $N=3$ in Fig.~\ref{fig:RepeaterNetwork}. The distillation rate without the quantum repeater $D^\text{no-repeater}_{\text{GHZ}, N}$ can be obtained by using Theorem~\ref{thm:Complete}. Therefore, we focus on the distillation rate with the quantum repeater $D^\text{repeater}_{\text{GHZ}, N}$. We first derive an upper bound on this rate. To this end, we combine a party $A_1$ and the repeater $R$ as a group. This grouping enables joint operations between $A_1$ and $R$. Therefore, a GHZ-state distillation rate after the grouping is at least as high as that before the grouping. Here, we calculate the MEB $M_{\mathcal{P}^*, \infty}(\psi_\text{in})$ with respect to the fixed partition $\mathcal{P}^* = \{ A_1R, A_2, \ldots, A_N  \}$ where $\psi_\text{in}$ denotes the input state on the $N+1$ nodes, viewed as an $N$-partite state after grouping $A_1$ and $R$. The number of Bell states between the group $A_1 R$ and the other parties is $(N-1)(a+b)$. Also, the number of Bell states between $A_i$ ($i = 2, \ldots, N$) and the other parties $(N-1)a + b$. Therefore, we have $S(A_1 R) = (N-1)(a+b)$ and $S(A_i) = (N-1)a + b$. When we trace out the group $A_1R$, the number of remaining Bell states is $\binom{N-1}{2} a$. In addition, the number of remaining Bell states after tracing out $A_i$ ($i = 2, \ldots, N$) is $(N-2)(a+b) + \binom{N-2}{2}a = \binom{N-1}{2}a + (N-2)b$, Therefore, we have $E_{\bar{1R}} = \binom{N-1}{2} a$ and $E_{\bar{i}} = \binom{N-1}{2}a + (N-2)b$. Hence, for the input $N$-partite complete graph, the MEB becomes
\begin{equation}
        \begin{split}
            M_{\mathcal{P}^*, \infty} (\psi_\text{in}) &= \frac{1}{N} \{ (N-1)(a+b) + (N-1)((N-1)a + b) \}\\
            &- \frac{1}{N(N-1)} \{ \binom{N-1}{2} a + (N-1)(\binom{N-1}{2} a + (N-2) b)  \}\\
            &= \frac{N}{2}a + b.
        \end{split}
\end{equation}
Then, using Theorem~\ref{Theorem:Upperbound} and the fact that the MEB is upper bounded by the MEB with respect to any fixed partition, we obtain
\begin{equation}
    D_{\text{GHZ},N}^\text{repeater} \leq D_{\text{GHZ}, N} (\psi_\text{in})  \leq M_{\mathcal{P}^*, \infty}(\psi_\text{in}) = \frac{N}{2}a + b.
\end{equation}

We next derive a lower bound on $D^\text{repeater}_{\text{GHZ}, N}$. To this end, we consider the following LOCC protocol. First, we use the original complete graph $K_N$ to generate GHZ states. Since $K_N$ has $\binom{N}{2}$ edges and one spanning tree requires $N-1$ edges, by performing the fractional spanning-tree packing, we can generate $\binom{N}{2}a / (N-1) = \frac{N}{2}a$ copies of GHZ states from $K_N$. Next, we consider the edges between the network repeater $R$ and the $N$ parties. When $R$ performs local operations on its qubits and the $N$ parties perform some local operations depending on the results of $R$'s local operations, the parties can share one copy of the GHZ state deterministically. In total, $\frac{N}{2}a + b$ copies of the GHZ state can be generated, i.e., $D^\text{repeater}_{\text{GHZ}, N} \geq \frac{N}{2}a + b$. Consequently, we obtain
\begin{equation}
        D^\text{repeater}_{\text{GHZ}, N} = \frac{N}{2}a + b.
\end{equation}

\subsection{2. Bipartition bounds}
In this subsection, we consider bipartition bounds on the GHZ-state distillation rates without and with the network quantum repeater. We start with the bipartition bound with the repeater. We split $N+1$ nodes into two groups $S$ and $\bar{S}$. Let $k$ denote the number of parties in the group $S$ for $1 \leq k \leq N-1$. The number of parties in $\bar{S}$ becomes $N-k$. Then, the number of edges between parties in $S$ and parties in $\bar{S}$ can be expressed as $k \times (N-k)$. When the network repeater $R$ belongs to $S$, the number of Bell states between $S$ and $\bar{S}$ is $k(N-k)a + (N-k)b$. On the other hand, when $R$ belongs to $\bar{S}$, the number of Bell states becomes $k(N-k)a + kb$. Therefore, we have the following upper bound on the GHZ-distillation rate:
\begin{equation}
    D^\text{repeater}_{\text{GHZ}, N} \leq \min_{1 \leq k \leq N-1} \{ k (N-k)a + \min \{ N-k, k \}b  \}.
\end{equation}
For arbitrary $1 \leq k \leq N-1$, we have $k(N-k) - (N-1) = (k-1) (N-k-1) \geq 0$ and $\min\{N-k, k  \} \geq 1$. Consequently, $k(N-k)a + \min\{ N-k, k \}b \geq (N-1)a + b$.
The equality holds when $k=1$ or $k = N-1$. Hence, the bipartition bound on the GHZ-distillation rate with the network quantum repeater is
\begin{equation}
    D^\text{repeater}_{\text{GHZ}, N} \leq (N-1)a + b.
\end{equation}
We can obtain the bipartition bound without the repeater by substituting $b = 0$ into the above bound with the repeater, i.e.,
\begin{equation}
    D^\text{no-repeater}_{\text{GHZ}, N} \leq (N-1)a.
\end{equation}

\section{H. Simultaneous distribution of GHZ and Bell states}
\subsection{1. Proof of Theorem~\ref{thm:Complete}}
Here, we describe the proof of Theorem~\ref{thm:Complete}. Consider a fixed partition $\mathcal{P}_0 = \{A_1, A_2, \ldots, A_N  \}$. We first calculate $M_{\mathcal{P}_0, \infty} (\Phi_{K_N})$. Since each party shares $N-1$ copies of the Bell state with the other parties, the von Neumann entropy of party $A_i$ is $S(A_i)_{\Phi_{K_N}} = N-1$. On the other hand, when we trace out the system of party $A_i$, the number of remaining Bell states becomes $\binom{N-1}{2}$. Then, using Lemma~\ref{lem:REEonBellPairNetwork} gives $E_{\bar{i}} (\Phi_{K_N}) = \binom{N-1}{2}$. Therefore,
\begin{equation}
        \begin{split}
            M_{\mathcal{P}_0, \infty} (\Phi_{K_N}) = \frac{1}{N}\sum_i S(A_i)_{\Phi_{K_N}} - \frac{1}{N(N-1)} \sum_i E_{\bar{i}}^\infty (\Phi_{K_N})= \frac{N}{2}.
        \end{split}
\end{equation}
Next, we calculate $M_{\mathcal{P}_0, \infty} (T_n)$. For the output state $T_n$, the von Neumann entropy of party $A_i$ becomes $S(A_i)_{T_n} = m_n + (N-1) k_n$. When we trace out $A_i$, the GHZ states become fully-separable states and the number of remaining Bell pairs is $\binom{N-1}{2} k_n$. Therefore, by using Lemma~\ref{lem:REEonBellPairNetwork} again, we have $E_{\bar{i}} (T_n) = \binom{N-1}{2} k_n$, and consequently
\begin{equation}
        \begin{split}
            M_{\mathcal{P}_0, \infty} (T_n) = m_n + \frac{N}{2}k_n.
        \end{split}
\end{equation}
Suppose an LOCC protocol $\Lambda_n$ transforms $n$ copies of $\Phi_{K_N}$ into a quantum state $\omega_n$ satisfying $T(\omega_n, T_n) \leq \epsilon_n$, where $\epsilon_n \to 0$ as $n \to \infty$. Since $T_n$ is a pure state, we can apply Proposition~\ref{prop:MEBhatExplicitLowerBound}. Here, the local dimension of $T_n$ is given by $d_n = 2^{m_n + (N-1)k_n}$. Therefore, by the fixed-partition version of Proposition~\ref{Prop:MEBhatLOCCMonotone} and Eq.~\eqref{eq:MEBhatExplicitLowerBoundFixed}, we obtain
\begin{equation}
    \begin{split}
        \hat{M}_{\mathcal{P}_0, \infty}(\Phi_{K_N}^{\otimes n}) &= \frac{N}{2}n \\
        &\geq M_{\mathcal{P}_0, \infty} (T_n) -\beta_N\sqrt{\epsilon_n}\log_2d_n -b_N(\sqrt{\epsilon_n})\\
        &=m_n+\frac N2k_n
-\beta_N\sqrt{\epsilon_n}[m_n+(N-1)k_n]
-b_N(\sqrt{\epsilon_n})
    \end{split}
\end{equation}
Dividing both sides by $n$ and taking $n \to \infty$, the error terms vanish because $\epsilon_n \to 0$ and the distillation rates $m_n/n$ and $k_n/n$ are finite. Hence, we obtain
\begin{equation}
        \begin{split}
            g + \frac{N}{2}r &\leq \frac{N}{2}.
        \end{split}
\end{equation}

We discuss the achievability of the above upper bound. To this end, we consider the following LOCC protocol. For $n$ copies of the Bell state on each edge, we keep $n \times r$ copies and use $n-n \times r$ copies for the fractional spanning-tree packing. The complete graph $K_N$ contains $\binom{N}{2} = N(N-1)/2$ edges and one spanning-tree requires $N-1$ edges. By Theorem~\ref{Theorem:BellpairNetwork}, the optimal GHZ-distillation rate of the complete graph $K_N$ is $N/2$. Then, the number of generated GHZ states from $n$ copies of the input state in this protocol is $\frac{N}{2} (1 -r)n$. Since the number of Bell states on each edge is $nr$, the simultaneous distillation rates of this LOCC protocol are
\begin{equation}
        g + \frac{N}{2}r = \frac{N}{2} (1-r) + \frac{N}{2} r = \frac{N}{2},
\end{equation}
that is, the above upper bound is achievable.

\subsection{2. Proof of Theorem~\ref{thm:Asymmetric}}
In this subsection, we describe the proof of Theorem~\ref{thm:Asymmetric}. First, we consider the bipartition $A:BC$. In this scenario, since $A$'s reduced state for one copy of the GHZ or the Bell state gives one bit of the von Neumann entropy, we have the following for the input state $\psi_{ABC}$, $S(A)_\psi = a + b$, and, for the final state $T_n$, $S(A)_{T_n} = m_n + k_{AB, n} + k_{AC, n}$. Let $\omega_n$ be the output of an LOCC protocol acting on $\psi_{ABC}^{\otimes n}$, and suppose that $T(\omega_n, T_n) \leq \epsilon_n$, where $\epsilon_n \to 0$. Then, the LOCC monotonicity (Eq.~\eqref{eq:REELOCCmonotone}) and the continuity (Eq.~\eqref{eq:REEContinuity}) of the REE give
\begin{equation}
    n(a+b) = E_R(A^n:B^nC^n)_{\psi^{\otimes n}} \geq E_R(A:BC)_{\omega_n} \geq E_R(A:BC)_{T_n} - o (n).
\end{equation}
Dividing by $n$ and taking $n \to \infty$ gives $g+x+y \leq a +b$. Analogously, considering the bipartitions $B:AC$ and $C:AB$ gives $g+x+z \leq a + c$ and $g + y + z \leq b + c$, respectively.

Consider a fixed partition $\mathcal{P}_0 = \{ A, B, C\}$. We next calculate $M_{\mathcal{P}_0, \infty}(\psi)$ for the input state $\psi_{ABC}$. The von Neumann entropy of each party is $S(A)_\psi = a+b$, $S(B)_\psi = a+c$, and $S(C)_\psi = b+c$. The regularized REE of the marginals on $AB$, $AC$, and $BC$ can be calculated as $E_R^\infty(A:B)_\psi = a$, $E_R^\infty(A:C)_\psi = b$, and $E_R^\infty(B:C)_\psi = c$, respectively. Hence $M_{\mathcal{P}_0, \infty} (\psi) = (a+b+c)/2$. Then, we calculate $M_{\mathcal{P}_0, \infty}(T_n)$. First, for the von Neumann entropy of each party, we have $S(A)_{T_n} = m_n + k_{AB, n}+k_{AC,n}$, $S(B)_{T_n} = m_n + k_{AB, n}+k_{BC,n}$, and $S(C)_{T_n} = m_n + k_{AC, n}+k_{BC,n}$. We consider a reduced state between $A$ and $B$. This state can be expressed as $(T_n)_{AB} = \Phi_{AB}^{\otimes k_{AB,n}} \otimes \sigma_{AB}^\text{SEP}$, where $\sigma_{AB}^\text{SEP}$ denotes a separable state between $A$ and $B$. Since tracing out $\sigma_{AB}^\text{SEP}$ is LOCC and the regularized REE is LOCC monotone (Eq.~\eqref{eq:RegularizedREELOCCmonotone}),
\begin{equation}
        E_R^\infty ((T_n)_{AB}) \geq E_R^\infty (\Phi_{AB}^{\otimes k_{AB, n}}) = k_{AB, n} E_R^\infty (\Phi_{AB}) = k_{AB, n},
\end{equation}
where we use the weak additivity of the regularized REE (Eq.~\eqref{eq:RegularizedREEWeakAdditivity}) for the first equality and we use Eq.~\eqref{eq:BipartiteREEonPureState}. On the other hand, due to the subadditivity of the REE (Eq.~\eqref{eq:REESubadditivity}),
    \begin{equation}
        \begin{split}
            E_R ((T_n)_{AB}) &\leq E_R (\Phi_{AB}^{\otimes k_{AB, n}}) + E_R (\sigma_{AB}^\text{SEP})= k_{AB, n},
        \end{split}
    \end{equation}
where the equality holds since the REE of separable states is zero. Moreover, the subadditivity of the REE (Eq.~\eqref{eq:REESubadditivity}) gives
    \begin{equation}
        \begin{split}
            E_R^\infty ((T_n)_{AB}) &= \lim_{l \rightarrow \infty} \frac{1}{l} E_R ((T_n)_{AB}^{\otimes l})\leq E_R ((T_n)_{AB})= k_{AB, n}.
        \end{split}
    \end{equation}
Consequently, $E_R^\infty ((T_n)_{AB}) = k_{AB, n}$. Similarly, we have $E_R^\infty ((T_n)_{AC}) = k_{AC, n}$ and $E_R^\infty ((T_n)_{BC}) = k_{BC, n}$. Therefore, $M_{\mathcal{P}_0, \infty} (T_n) = m_n + (k_{AB, n} + k_{AC, n} + k_{BC, n})/2$. Since the input state is pure, $\hat{M}_{\mathcal{P}_0, \infty} (\psi^{\otimes n}) = n M_{\mathcal{P}_0, \infty}(\psi)$. Then, applying the fixed-partition version of Proposition~\ref{Prop:MEBhatLOCCMonotone} to $\omega_n$ and then applying Eq~\eqref{eq:MEBhatExplicitLowerBoundFixed} gives
\begin{equation}
    n M_{\mathcal{P}_0, \infty}(\psi) \geq \hat{M}_{\mathcal{P}_0, \infty}(\omega_n) \geq M_{\mathcal{P}_0, \infty} (T_n) - \beta_3 \sqrt{\epsilon_n} \log_2 d_n - b_3 (\sqrt{\epsilon_n}),
\end{equation}
where $d_n$ is the maximal local dimension of $T_n$ and $\epsilon_n \rightarrow 0$ as $n \rightarrow \infty$. Since $\log_2 d_n = O(n)$, we have
\begin{equation}
        m_n + \frac{1}{2} (k_{AB, n} + k_{AC, n} + k_{BC, n}) \leq \frac{a+b+c}{2}n + o(n).
\end{equation}
Dividing both sides by $n$ and taking $n \rightarrow \infty$ gives $2g + x + y + z \leq a + b +c$. 
    
We consider the achievability of Eq.~\eqref{eq:thmAsymmetric}. Define $d_a \coloneqq a - x$, $d_b \coloneqq b - y$, and $d_c \coloneqq c-z$. Then, we can rewrite Eq.~\eqref{eq:thmAsymmetric} as
\begin{equation}\label{eq:SimultaneousDistributionAsymmetricD}
\begin{split}
            d_a + d_b + d_c &\geq 2 g, \quad d_a + d_b \geq g, \quad d_a + d_c \geq g, \quad d_b + d_c \geq g.
\end{split}
\end{equation}
We first consider the case where $d_a, d_b, d_c \geq 0$. In this case, $x$ copies of $\Phi_{AB}$, $y$ copies of $\Phi_{AC}$, and $z$ copies of $\Phi_{BC}$ are kept and the remaining Bell states are used for generating GHZ states. In the triangle network, there are three spanning trees and let $t_a, t_b,$ and $t_c$ denote GHZ-distillation rates using the edges $AC$ and $BC$, $AB$ and $BC$, and $AB$ and $AC$, respectively. These rates must satisfy $t_a + t_b + t_c = g$.
Moreover, the rates must satisfy $g - t_a \leq d_a$, $g- t_b \leq d_b$ and $g - t_c \leq d_c$. A necessary and sufficient condition that there exist rates $t_a, t_b$ and $t_c$ satisfying the above conditions is
\begin{equation}\label{eq:NecessarySufficient}
        (g - d_a)_+ + (g-d_b)_+ + (g-d_c)_+ \leq g,
\end{equation}
where $(x)_+ = \max{\{x, 0\}}$. We show that $d_a, d_b$, and $d_c$ satisfying Eq.~\eqref{eq:SimultaneousDistributionAsymmetricD} satisfy the necessary and sufficient condition as follows. When the number of $d_i < g$ ($i = a, b, c$) is zero, the left-hand side of Eq.~\eqref{eq:NecessarySufficient} becomes zero and the inequality holds. When the number of $d_i < g$ is one, the inequality holds since $d_i \geq 0$. When $d_i, d_j < g$ ($i \neq j$), since $d_i + d_j \geq g$ from Eq.~\eqref{eq:SimultaneousDistributionAsymmetricD}, we have $(g- d_i)_+ + (g-d_j)_+ = 2g - (d_i + d_j) \leq g$. Finally, when $d_a, d_b, d_c < g$, using $d_a + d_b + d_c \geq 2g$ gives $(g- d_a)_+ + (g-d_b)_+ + (g-d_c)_+ = 3g - (d_a+d_b+d_c) \leq g$. In conclusion, for arbitrary $d_a, d_b, d_c \geq 0$ satisfying Eq.~\eqref{eq:SimultaneousDistributionAsymmetricD}, the necessary and sufficient condition in Eq.~\eqref{eq:NecessarySufficient} holds. That is, there exists an LOCC protocol whose distillation rates satisfy Eq.~\eqref{eq:SimultaneousDistributionAsymmetricD}.

Next, we consider the case where $d_a < 0$ and $d_b, d_c \geq 0$. For $u > 0$, let $d_a = -u$. Since $d_a + d_b \geq g$ and $d_a + d_c \geq g$ from Eq.~\eqref{eq:SimultaneousDistributionAsymmetricD}, we have $d_b = b - y \geq g - d_a = g + u$ and $d_c = c-z \geq g + u$. Here, we consider the following LOCC protocol. First, the parties keep $a$ copies of $\Phi_{AB}$, $y$ copies of $\Phi_{AC}$, and $z$ copies of $\Phi_{BC}$. Then, the parties use $u$ copies of $\Phi_{AC}$ and $\Phi_{BC}$, and generate $u$ copies of $\Phi_{AB}$ by entanglement swapping. Finally, using $g$ copies of $\Phi_{AC}$ and $\Phi_{BC}$, the parties generate $g$ copies of the GHZ state. We can consider similar LOCC protocols for the cases where only $d_b$ or $d_c$ is negative. Here, at most one of $d_a, d_b, d_c$ can be negative because, when $d_i, d_j < 0$ for $i \neq j$, the condition $d_i + d_j \geq g$ in Eq.~\eqref{eq:SimultaneousDistributionAsymmetricD} does not hold. Therefore, for arbitrary distillation rates $g, x, y, z$ satisfying Eq.~\eqref{eq:thmAsymmetric}, there exist concrete LOCC protocols achieving the rates, concluding the proof.

Theorem~\ref{thm:Asymmetric} focuses on the input state expressed as a product of Bell states. However, we can generalize the result to products of arbitrary bipartite pure states straightforwardly:
\begin{cor}[]
    Let an input state for each round be $\ket{\psi_\text{in}} = \ket{\phi_{AB}} \otimes \ket{\phi_{AC}} \otimes \ket{\phi_{BC}}$ where $\ket{\phi}_{AB}$, $\ket{\phi}_{AC}$, and $\ket{\phi}_{BC}$ are arbitrary pure states. Let $a = S(\text{Tr}_B[\phi_{AB}])$, $b = S(\text{Tr}_C[\phi_{AC}])$, and $c = S(\text{Tr}_C[\phi_{BC}])$.
    Then, the simultaneous distillation rates of GHZ and Bell states satisfy
    \begin{equation}
        \begin{split}
            2g + x + y+ z \leq a + b + c, \quad g + x + y \leq a + b, \quad g + x + z \leq a + c, \quad g + y + z \leq b + c.
        \end{split}
    \end{equation}
\end{cor}

\begin{proof}
    For bipartite pure states, the optimal rates of entanglement concentration and dilution are both given by the entropy of entanglement~\cite{Bennett1996Concentrating}. Then, in the asymptotic limit, the input state $\psi_\text{in}^{\otimes n}$ can be prepared from Bell states shared between $AB$, $AC$, and $BC$ at rates $a$, $b$, and $c$, respectively, with vanishing error by entanglement dilution. Therefore, the bounds in Theorem~\ref{thm:Asymmetric} also apply to the original pure-state network. On the other hand, the input state can be converted into Bell states at the same asymptotic rates by entanglement concentration. Then, by Theorem~\ref{thm:Asymmetric}, the bounds are achievable, concluding the proof.
\end{proof}

\subsection{3. Bipartition bounds}
The bipartition bound on GHZ and Bell-state distillation in the triangle quantum network with the input state $\psi_{ABC} = \Phi_{AB}^{\otimes a} \otimes \Phi_{AC}^{\otimes b} \otimes \Phi_{BC}^{\otimes c}$ and the target state $T_n = (\Gamma_2^{(3)})^{\otimes m_n} \otimes \Phi_{AB}^{\otimes k_{AB,n}} \otimes \Phi_{AC}^{\otimes k_{AC, n}} \otimes \Phi_{BC}^{\otimes k_{BC, n}}$ is given by
\begin{equation}
    g+x+y \leq a + b, \quad g + x + z \leq a + c, \quad g + y + z \leq b + c.
\end{equation}
The proof of this bound is the same as the proof of Theorem~\ref{thm:Asymmetric}. In addition, we can straightforwardly extend the proof to derive the bipartition bound in the $N$-partite complete graph $K_N$ with equal Bell-state distillation rates $r$:
\begin{equation}
    g + (N-1)r \leq N-1.
\end{equation}

\section*{Supplementary References}

\putbib
\end{bibunit}

\end{document}